\documentclass[a4paper,10pt]{article}

\usepackage{lscape}
\usepackage{cancel}
\usepackage{bm}
\usepackage{bbm}
\usepackage{multirow} 
\usepackage{amsfonts}
\usepackage{amssymb}
\usepackage{graphicx}
\usepackage{amsmath}
\usepackage[usenames,dvipsnames]{color}
\usepackage{hyperref}
\usepackage[left=1in,right=1in,top=1in,bottom=1in]{geometry}             

\hypersetup{
    colorlinks=true,         
    linkcolor=blue,          
    citecolor=red,        
    urlcolor=Violet             
}

\newcommand{\st}[1]{\textrm{\tiny{#1}}}

\def\be{\begin{equation}}
\def\ee{\end{equation}}
\def\bea{\begin{eqnarray}}
\def\eea{\end{eqnarray}}

\title{Back to Parmenides}
\author{\bf Henrique Gomes\footnote{\href{mailto:gomes.ha@gmail.com}{gomes.ha@gmail.com}}\\\it Perimeter Institute for Theoretical Physics\\ \it 31 Caroline Street, ON, N2L 2Y5, Canada}
\begin{document}
\maketitle
\begin{abstract}
After a brief  introduction to issues that plague the realization of a theory of quantum gravity, I suggest that the main one concerns a quantization of the principle of relative simultaneity. This leads me to a distinction between time and space, to a further degree than that present in the canonical approach to general relativity. With this distinction, one can make sense of superpositions as interference between alternative paths in the relational configuration space of the entire Universe. But the full use of relationalism brings us to a timeless picture of Nature, as it does in the canonical approach (which culminates in the Wheeler-DeWitt equation). After a discussion of Parmenides and the Eleatics'  rejection of time, I show that there is middle ground between their view of absolute  timelessness and a view of physics taking place in timeless configuration space. In this middle ground, even though change does not fundamentally exist, the illusion of change can be recovered in a way not permitted by Parmenides. It is recovered  through a particular density distribution over configuration space which gives rise to \lq{}records\rq{}. Incidentally, this distribution seems to have the potential to dissolve further aspects of the measurement problem that can still be argued to haunt the application of decoherence to Many-Worlds quantum mechanics. I end with a discussion indicating that the conflict between the conclusions of this paper and our  view of the continuity of the self  may still intuitively bother us. Nonetheless, those conclusions should be no more challenging to our intuition than  Derek Parfit\rq{}s   thought experiments on the subject. 
   \end{abstract}

\tableofcontents

\section{ A summary of the construction (for the experts).}
Quantum mechanics arose in the 1920\rq{}s. General relativity has been around since the 1910\rq{}s. But, as of 2018, we still have no quantum theory of the gravitational field. What is taking us so long? I believe the most challenging obstacle in our way is understanding the quantum superposition of general relativistic causal structures. This obstacle is couched on facets of the `problem of time' \cite{Kuchar_time} --- an inherent difficulty in reconciling a picture of time evolution in quantum mechanics to a \lq{}block time\rq{} picture of general relativity. 

I also believe  we can overcome this obstacle only if we accept a fundamental distinction between time and space. The distinction is timid in general relativity -- even in its ADM  form  \cite{ADM} -- and here I want to push it further.  In this spirit,  I will consider space to be fundamental and time to be a derived concept -- a concept at which we arrive from  change (a loose quote from Ernst Mach).  

 I will here investigate the consequences of this distinction between time and space. The distinction only allows a restricted class of fundamental physical fields   --- the ones whose content is spatially relational --- and it thus also restricts the sort of fundamental theories of reality. It is consequential in that with this view we must reassess our interpretation of quantum mechanics and its relationship to gravity.  The new interpretation is compatible with a version of timelessness that I will explain below. With timelessness, comes the requirement of explaining history without fundamental underlying dynamics. The role of dynamics is fulfilled by what I define as \lq{}records\rq{}.
 
  This paper will consist mainly of two parts: one justifying the timeless approach through problems in quantum gravity, and another describing physics within a general timeless theory proposed here. The next subsection, \ref{sec:sym_rel}, gives  a brief look into the justification part to come, i.e. describing the need for different relational symmetries in the present framework. These reasons will be furthered in section \ref{sec:QG_problems}, and the symmetries explicitly presented in \ref{sec:symmetries}. In the next subsection, \ref{sec:timeless_intro}, I will give a brief overview of the second part of the paper, i.e. constructing physical theories which make sense in the timeless setting.

\subsection{Symmetries and relationalism}\label{sec:sym_rel}
Is a non-spacetime-covariant perspective  in conflict with what we have learned from classical general relativity? Not necessarily. For instance, Lorentz invariance can be a property recovered for  \emph{classical} solutions of the field equations  \cite{Steve_SD}.   From this point of view, Einstein's unification of space and time would be a property of  solutions of the equations of motion rather than a foundational principle of the theory.   In fact, there are many hints that by imposing space-time covariance quantum mechanically we might be overstepping the principle\rq{}s jurisdiction, and I will list them in section \ref{sec:QG_problems}.  

For now, following the ideas of Mach further, we find that general relativity falls short of his relational demands on Nature. According to Mach, the properties of objects can have no intrinsic meaning -- it is only by comparison that we obtain objective knowledge. For example, in Newtonian mechanics, only relative positions of particles should enter our description of the Universe. There is no objective meaning in the absolute location of a particle, only in its relation to all the other constituents of the Universe \cite{Barbour_Mach}. Similarly, scale could be argued to only be meaningful in terms of comparisons. No one has ever measured, nor will ever measure, a fundamental scaleful quantity. 

 Nonetheless, in general relativity, even in vacuum, the proper length of spacetime curves sets a scale,  irrespective of any ruler with which their lengths could be compared. In more mathematical jargon, the spacetime length of a (non-null) curve is not generally invariant under conformal transformations of the spacetime metric.\footnote{The lengths of \emph{null} curves are indeed invariant under conformal transformations. But  even without matter the Einstein-Hilbert action is not explicitly conformally invariant. Upon a conformal transformation the action gains a kinetic term for the conformal factor.}  

  If we accept the ontological split of spacetime into space and time, we prioritize a different kind of symmetry: those that act on each spatial field configuration. On the other hand, symmetries that had previously acted naturally on spacetime lose their simplicity and their privileged status. These notions will be expanded in section \ref{sec:symmetries} and used to guide our search for  symmetries more sympathetic to quantum evolution. 
  
   Focusing on space and its symmetries rather than spacetime and its symmetries, our postulates demand complete relationalism, to the satisfaction of Mach\rq{}s  demands. 
Then we find that there are  two types of symmetries -- of scale and of position -- implied by spatial relationalism. Thus both the local scale and the spatial position are to be defined  relationally -- that is without ever referring to any absolute frame of reference. 
These two types of transformation act \lq{}intrinsically\rq{}   on each spatial metric configuration. By \lq{}intrinsic\rq{}, I mean that their action on a configuration should only depend on the configuration itself,  not on other configurations (even if neighboring ones), which is why I call  it a \lq{}law of the instant\rq{}. 

In the quantum mechanical context, symmetries compatible with a timeless transition amplitude should also be \lq{}laws of the instant\rq{}.    Fortuitously, at least in metric variables the two relational symmetries  mentioned above represent the most general \lq{}laws of the instant\rq{}.  Identifying configurations related by such symmetries, the  physical configuration space of a timeless theory becomes a quotient space, and more, a \lq{}principal fiber bundle\rq{} \cite{Bleecker}. In the case of gravity, this quotient is called conformal superspace. A curve in conformal superspace describes changes of scale-free geometry -- or shapes -- as it goes along.

On the other hand, symmetries that are inherited from a spacetime picture need not be laws of the instant.  For example, the usual symmetry of refoliations is a remnant  of spacetime diffeomorphisms. It is not a \lq{}law of the instant\rq{} and is thus inimical to spatial relationalism as I define it here.\footnote{It is not enough to note that refoliations act differently on phase space than  spatial diffeomorphisms (e.g. act as a groupoid vs  as a group). Both transformations have infinitesimal actions on phase space points after all, and, apart from matters having to do with chaos \cite{Philipp_chaos}, it is not clear why the action of one would permit a quotienting procedure, but not the other. Here the difference is made explicit. } 

There is one unintended consequence of ridding ourselves of symmetries that move time: how do we regain time evolution? A Newtonian view of time seems outdated and stale. After the broader lessons of general relativity regarding relationalism, ``time, of itself, and from its own nature, flowing equably without regard to anything external'', is unconvincing. With full relationalism must come timelessness. 


\subsection{Timelessness}\label{sec:timeless_intro}
The ideas developed here have ancient roots. They are a modern version of the classical debate between Parmenides and Heraclitus. Heraclitus held  that nothing in this world is constant except change and becoming. Parmenides on the other hand believed that the ontic changes  we experience with our senses are deceptive;  behind the veil of our perceptions lies the changeless true nature of the World.

  A much more recent  incarnation of the Parmenidean view is found in the work of Julian Barbour (see \cite{Julian_End}), which I use here as a nearer port of departure. Barbour observes that timeless configuration space should be seen as  the realm containing every possible \lq{}now\rq{},  or instantaneous configuration of the universe. In \cite{Julian_QGII}, Barbour attempts to accommodate timelessness more intuitively into our experience:
  \begin{quote}
  An alternative is that our direct experience, including that
of seeing motion, is correlated with only \emph{configuration} in our brains: the correlate of the
conscious instant is part of a point of configuration space [...] Our seeing motion at some instant is correlated with a single
configuration of our brain that contains, so to speak, several stills of a movie that we are
aware of at once and interpret as motion. [...]
  Time is not a framework in which the configurations of the world
evolve. Time exists only so far as concrete configurations express it in their structure. The
instant is not in time; time is in the instant.
  \end{quote}
I \textit{almost} wholeheartedly agree with Barbour. We diverge only in the attribution of \textit{experience} to \textit{each} configuration alone. I believe there is no empirical access to single field configurations, therefore all statements about experience refer to some coarse-graining, or regions of configuration space, where a given attribute is represented. Clearly, this empirical dilution of Barbour\rq{}s radical \lq{}solipsism of the instant\rq{} does not conflict with my theoretical reasons for considering instantaneous configuration space to be ontologically fundamental.

\subsubsection{Timelessness and the relational quantum transition amplitude}
When there is some underlying notion of the passage of time, we think about configuration $q_1$ (and possibly some neighborhood thereof)  as `now'; it is a configuration which has evolved from earlier initial conditions (e.g.: the Big Bang). We think of $q_2$ as some configuration in the future, and we want to predict the probability that our Universe will somehow become $q_2$. 
 Conversely, we can think of this amplitude as saying that, having measured $q_1$ at a previous time, closing our eyes, letting some time elapse, we now want to calculate the probability that when we open our eyes we will find ourselves in $q_2$. Both alternatives represent the probability of $q_2$ \lq{}conditional\rq{} on $q_1$, but I believe the latter is more representative of how we actually use predictions.   

  In section \ref{canonical_timeless}, I will introduce  the transition amplitude $W(q_1,q_2)$, which in the timeless context   is the substitute of $G((t_1,q_1),(t_2,q_2))$ --- the quantum transition  amplitude between an initial and final configuration when an absolute notion of time exists.  Such amplitudes have been studied in the path integral language by  Chiou \cite{Chiou}. According to him,  $W(q_1,q_2)$ represents `the amplitude related to measuring $q_2$, given that  $q_1$ was measured at a \emph{previous instance}\rq{}.   

  Although $W(q_1,q_2)$ appears to be mathematically self-consistent, we are still left with a host of questions unanswered. What `previous instance\rq{}?  How do we know  that the previous configuration really `happened\rq{}, and how did it \lq{}become\rq{} the present one? When Time is to be a derived concept at which we arrive from change, what is it that promotes change itself?  What is it that drives the present to become the future? Indeed, most relational approaches to quantum mechanics based on the transition amplitude do not address this question \cite{Rovelli_book}.


  Again quoting Barbour \cite{Julian_QGII} as our proxy for Parmenides
 \begin{quote}
 By analogy with Descartes\rq{}s Cogito ergo sum, we know that the present instant is
actualized. However, because we can never step out of the present instant, we can never
know if any other instant is actually experienced. For we shall never know whether other possible instants, including
what we take to be our own past, are actual or whether the present instant is unique. \end{quote}   
     Indeed, it seems undeniable that the present -- not just the $t=$now, but each and every present -- has a preferred ontological status compared to either its past or its future. Perhaps the difficulty of assigning ontological meaning to $W(q_1,q_2)$ is a reflection of this fact. 
     
      My answer will be that the past doesn\rq{}t become the present -- it is only embedded in the present. Every present exists, every present is unique, and some presents  may be entangled with other presents. 
  
 Everything we experience at any instant, including  memories, must be etched into our instantaneous brain configuration. Records of apparent past events should exist in the present configuration and its neighbors. This is what Barbour alludes to by saying that \lq{}time is in the instant\rq{}. But why do records give  a consistent picture of the past? At any moment we are in the possession of a host of redundant records of the same event, and they had better be in mutual accord. It is from this consistent mosaic of records that we build models of the laws of Nature. We  fit the pieces together into a larger explanatory framework we call \lq\lq{}science\rq\rq{}.

 \subsubsection{Timeless records and the volume form.}  
 
 After a description of timeless quantum mechanics and its path integral formulation in section \ref{canonical_timeless}, I  will introduce a volume form -- or probability density -- in configuration space, and a notion of semi-classical records -- the key constructions to answer  these questions. Roughly, the volume-form represents a counting of possible instantaneous Universes.

    Since certain spatial field configurations may  also contain an observer\rq{}s conscious state, the volume element around such a configuration expresses the amount of observers which all possess  similar experiences.    The volume form will depend on the given Lagrangian action, and we can determine paths that extremize this action between two configurations (if any such path  exists). 
    
 The kernel of the argument lies in accepting timelessness  by interpreting $q_2$ as configuration \lq{}now\rq{},  but having $q_2$ contain \emph{records} of $q_1$.\footnote{To me, this is related to what Barbour has called \lq{}time capsules\rq{}, but Barbour has objected to this connection.}  In other words, even if there is no concrete sense in which $q_1$  \lq{}previously happened\rq{}, in the presence of records of $q_1$, the volume of configurations in the region around $q_2$ become tightly correlated with the volume of configurations around $q_1$.  The connection between the two infinitesimal volume forms  implies that if there aren\rq{}t many copies of an observer close to $q_1$, there will be even less around $q_2$.  This not only guarantees a certain notion of unitarity, but, as a conditional probability, can be interpreted as an inference of the existence of the past:  $q_1$ can be inferred from $q_2$. 

    Furthermore  I can show that:  If $q_2$ has a record of $q_1$ \emph{and} there is a unique extremal path between the two configurations -- the classical limit-- then the entire path has at least a granular ordering of records. Namely,  parametrizing the path, $\gamma(t)$,  such that $\gamma(0)=q_1\,,\gamma(t^*)=q_2$, then $\gamma(t\rq{}\rq{})$ has a record of $\gamma(t\rq{})$ iff $t\rq{}<t\rq{}\rq{}$ and $|t\rq{}\rq{}-t\rq{}|>> t_{\mbox{\tiny planck}}$. 
      
     In the classical limit, we are led to the idea of a complete past history, culminating always in the present. Different lines of evidence for the objective existence of the record $q_1$ seem to concur. There are redundant records of the breakfast donut -- both my memory of it and the jelly stain on my shirt. After all, in this limit every aspect of our present configuration $q_2$ is connected to $q_1$ by a classical history of the fields.  This nesting of records is also what allows us to effectively use \lq\lq{}repeatability\rq\rq{} of experiments. Even though we only have access to \lq{}now\rq{},  each previous run of an experiment can be encoded in a future run,  in a very redundant manner subject to checks for consistency \cite{Riedel}.  It allows  frequencies to come into science. 
 
  From a Bayesian perspective, we are in fact completely  justified in assuming that \emph{configuration $q_2$ evolved from $q_1$} in the classical limit, realizing every configuration in between. The reconstruction of the past is complete, at least as a working hypothesis. 
  
  \subsubsection{Interference, not superposition.}
 Once spacetime is nothing but the relational change of space, superposing geometries becomes a more straightforward problem to tackle. In the path integral formalism, quantum mechanical superpositions of three-geometries can be realized through interfering paths in physical configuration space -- the space of all relationally equivalent three-geometries.

This is a counter-intuitive view on superpositions. We instinctively believe  quantum subsystems exist \emph{within} a spacetime, where they are joined by us, observers of these subsystems. Both subsystem and observer can independently be in states of superpositions. We naturally  expect therefore to see superpositions of subsystems within spacetime. But if instead we experience only instantaneous configurations -- and these include gravitational degrees of freedom --  we should expect something different.

 Let me explain what is it that we should expect in the relational context of a closed Universe. Each \emph{possible} `spacetime' (not necessarily of the generally-covariant sort)  is identified with a curve in physical configuration space  (configuration space modulo relational symmetries). In this context, the timeless path integral formulation of gravity resembles the usual timeless  path integral of quantum mechanics.

This analogy allows us to borrow from consistent histories \cite{CH_review} their notion of decoherence, or rather, lack thereof -- interference. In the path integral formulation of consistent histories in configuration space, one measures superposition through the use of a mathematical object called the `decoherence functional'.  Given an `in' and an `out' configuration, $q_1\,,q_2$, and a  partitioning of  paths between $q_1$ and $q_2$ (called a \emph{coarse-graining of the paths}), the decoherence functional gives a measure of interference between the elements of the partition. 

With a preferred notion of space, and the accompanying preferred choice of relational field-configuration space,  there is no need to wrap our minds around the meaning of `a superposition of spatial metrics'. Instead, certain types of configurations are experienced more, or less, frequently, according to how often they are represented in configuration space. I will clarify the meaning of this tricky word, \lq{}often\rq{}, in the sequel. 

 Nonetheless, this new theory brings with it new questions: about locality,  Bayesian probabilities, Born rule, etc,  which I tried to answer as fully as I could in \cite{Locality_riem, QG_deco, Conformal_geodesic} and which  I will only touch on here. 

In the following section, I will introduce more technical reasons for my interest in timeless theories. These have to do with quantum gravity. I thus start with a brief description of what would count as a theory of quantum gravity, before moving on to the sort of problems it has, which I believe timelessness might cure. In section \ref{canonical_timeless} I discuss timelessness in earnest.  I start with a little ancient history, centered on Parmenides and Zeno, who I think already grasped important aspects of the modern discussion. I then move on to expounding the standard timeless quantum path integral formulation, since it provides a useful tool for many of the results and intuitions to be developed. In section \ref{sec:timeless_QG}, I develop the configuration spaces for gravity in which this timeless formulation would be physically meaningful. Nevertheless, the formulation does not address the question of whether one would still require a \lq\lq{}driver\rq\rq{} of change. This driver becomes obsolete with the introduction of records, in section \ref{sec:records}. However,  records require on their turn a notion of a global \lq\lq{}initial state\rq\rq{}, or Past Hypothesis, which is introduced still in section \ref{sec:records}. In section  \ref{sec:Parfit}, I discuss psychological implications of this view on timelessness, and expand the mathematical discrepancy between this view and the Eleatic conclusions. I then conclude.

\section{The problems with quantizing gravity}
I will start in section  \ref{sec:tale}  with  what I believe are the main principles of quantum mechanics and gravity. I will then follow in section \ref{sec:problems} with a brief idiosyncratic exposition of in issues in quantizing gravity in section.  Then, in section \ref{sec:QG_problems}  I move on to an illustration  of which of these problems signal a fundamental difference between time and space. Finally, in section \ref{sec:symmetries}, I posit what sort of different fundamental symmetries --- i.e. to be imposed also off-shell, or at the quantum mechanical level --- would assuage the clash between dynamics and symmetry. These turn out to be the spatial relational symmetries. 

\subsection{A tale of two theories}\label{sec:tale}

General relativity  is one of the pillars of our modern understanding of the Universe, deserving a certain degree of familiarity from all those who purport to study Nature, whether from a philosophical or mathematical point of view. The theory has such pristine logical purity that it can be comprehensively summarized by John A. Wheeler's famous quip \cite{MTW}: 
\be\label{sentence:Wheeler}\mbox{\lq\lq{}Matter tells spacetime how to curve, and spacetime tells matter how to move.\rq\rq{}}\ee
 We should not forget however, that ensconced within Wheeler's sentence is our conception of spacetime as a dynamical geometrical arena of reality: no longer a fixed stage where physics unfolds, it is part and parcel of the play of existence. 

In mathematical terms, we have: 
\be\label{equ:GR} \underset{\mbox{spacetime curving}}{\underbrace{R_{\mu\nu}-\frac{1}{2}Rg_{\mu\nu}}}\propto\underset{\mbox{sources for curving}}{\underbrace{\phantom{\frac{1}{1}}T_{\mu\nu}}}
\ee
Given the sources, one will determine a geometry given by the spacetime metric $g_{\mu\nu}$ -- the ` matter tells spacetime how to curve'  bit. Conversely,  it can be shown that very light, very small particles will roughly follow geodesics defined by the geometry of the lhs of the equation  -- the `spacetime tells matter how to move'  part.\footnote{This distinction is not entirely accurate, as the rhs of  equation \eqref{equ:GR} usually  also contains the metric, and thus the equation should be seen as a constraint on which kind of space-times with which kind of matter distributions one can obtain, \lq\lq{}simultaneously\rq\rq{}. I.e. it should be seen as a spacetime, block universe, pattern, not as a causal relation (see \cite{Lehmkuhl}). }

  A mere decade after the birth of GR, along came quantum mechanics.  It was a framework that provided unprecedented accuracy in experimental confirmation,  predictions of new physical effects and a reliable compass for the construction of new theories. 
And yet, it has  resisted the intuitive understanding that was quickly achieved with general relativity. 
A much less accurate characterization than Wheeler's quip for general relativity has been borrowed from the pessimistic adage ``everything that can happen, does happen".\footnote{Recently made the title of a popular book on quantum mechanics \cite{Cox_book}.} The sentence is meant to raise the principle of superposition to the status of core concept of quantum mechanics (whether it expresses this clearly or not is very much debatable).   

In mathematical terms, the superposition principle can be seen in the Schr{\"o}dinger equation:
\be\label{equ:Schroedinger}
\hat{H}\psi=-i\hbar\frac{d}{dt}\psi
\ee
whose linearity implies that two solutions $\psi_1$ and $\psi_2$ add up to a solution $\psi_1+\psi_2$. In the path integral representation, superposition is built-in. The very formulation of the generating function is a sum over all possible field configurations $\phi$, 
\be\label{generating} \mathcal{Z}[j]=\int \mathcal{D}\phi\, \exp\left[{i\int \mathcal{L}(\phi)/\hbar}\right]\ee
where $\mathcal{L}(\phi)$ is the Lagrangian density for the field $\phi$ and $\mathcal{D}\phi$ signifies a summation over all possible values of this field (the second integral is over spacetime).   

Unfortunately, for the past 90 years, general relativity and quantum mechanics have not really gotten along. Quantum mechanics soon claimed a large chunk of territory in the theoretical physics landscape, leaving a small sliver of no-man's land also outside the domain of general relativity. 
In most regimes, the theories will stay out of each other's way -  domains of physics where both effects need to be taken into account for an accurate phenomenological description of Nature are hard to come by.   Nonetheless,  such a reconciliation might be necessary even for the self-consistency of general relativity: by predicting the formation of singularities,  general relativity ``predicts its own demise", to borrow again the words of John Wheeler. Unless, that is,  quantum effects can be suitably incorporated to save the day at such high curvature regimes.  
 
\subsection{The problems of quantum gravity}\label{sec:problems}
At an abstract level, the question we need to  face when trying to quantize general relativity is: how to write down a theory that includes all possible superpositions and yet yields something like equation \eqref{equ:GR} in appropriate classical regimes? Although  the incompatibility between general relativity and quantum mechanics can be of technical character, it is widely accepted that it has more conceptual roots.   In the following, I will describe only two such roots.

\subsubsection{Is non-renormalizability the only problem?}

The main technical obstacle cited in the literature is the issue of perturbative renormalizabilty. Gravity is a non-linear theory, which means that geometrical disturbances around a flat background can act as sources for the geometry itself. The problem is that unlike what is the case in other non-linear theories, the `charges'  carried by the non-linear terms in linearized general relativity become too `heavy' ---the gravitational coupling constant has negative mass dimensions--- generating a cascade of ever increasing types of interactions once one goes to high enough energies. This creates problems for treating such phenomena scientifically, for we would require an infinite amount of experiments to determine the strength of these infinite types of interactions. This problem can be called  \lq{}loss of  predictability\rq{}.

There are theories, such as Horava-Lifschitz gravity \cite{Horava}, which seem to be naively perturbatively renormalizable.  The source of renormalizability here is the greater number of spatial derivatives as compared to that of time derivatives. This imbalance  violates fundamental Lorentz invariance, breaking up spacetime into space and time.  Unfortunately, the theory introduces new degrees of freedom that appear to be problematic (i.e. their influence does not disappear at observable scales). 

And perhaps perturbative non-renormalizability is not the only problem. Indeed, for some time we have known that a certain theory of gravity called `conformal gravity' (or `Weyl squared')  is also perturbatively renormalizable. The problem there is that the theory is sick. Conformal gravity  is not a unitary theory, which roughly means that probabilities will not be conserved in time. But, which time? And is there a way to have better control over unitarity? This can be called \lq{}the problem of unitarity\rq{}.\footnote{Another approach to quantum gravity called Asymptotic Safety \cite{Reuter} also suffers from such a lack of control of unitarity. This approach also explores the possible existence of gravitational  theories whose renormalization will only generate dependence on a finite number of coupling constants, thus avoiding the loss of predictibility explained above.}   

 \subsubsection{A dynamical approach}

In covariant general relativity, the fundamental field, $g_{\mu\nu}$, already codifies causal relations, whether or not the equations of motion (the Einstein equations) have been imposed. 
So a first question to ask is if we can give a formulation of quantum gravity which reflects the fundamental distinction between causal and acausal.  One way of approaching this question is to first use a more dynamical account of the theory. We don't need to reinvent such an account -- it is already standard in the study of gravity, going by the acronym of ADM (Arnowitt-Deser-Misner) \cite{ADM}. The main idea behind a dynamical point of view is to set up initial conditions on a spatial manifold $M$ and construct the spacetime geometry by evolving in a given auxiliary definition of time.\footnote{ In this constructed space-time, the initial surface must be Cauchy, implying that one can only perform this analysis for space-times that are time-orientable.} Indeed most of the work in numerical general relativity requires the use of the dynamical approach. 
Such formulations allow us to use the tools of the Hamiltonian formalism of quantum mechanics to bear on the problem of quantizing gravity. With these tools,  matters regarding unitarity  are much easier to formulate, because there is a time with respect to which probabilities are to be conserved. 

However, since the slicing of spacetime  is merely an auxiliary structure, the theory comes with a constraint -- called the Hamiltonian constraint -- which implies a freedom in the choice of such artificial time slicings. The metric associated to each equal time slice, $g_{ab}$, and its associated momenta, $\pi^ {ab}$, must be related by the following relation at each spatial point $x\in M$: 
\be\label{equ:Hamiltonian} H(x):=R(x)-\frac{1}{g}\left(\pi^{ab}\pi_{ab}-\frac{1}{2}\pi^ 2\right)(x)=0\qquad \forall x\in M
\ee
where $g$ stands for the determinant of the metric.

\subsection{Problems of Time}\label{sec:QG_problems}

The constraints \eqref{equ:Hamiltonian} are commonly thought to guarantee that observables of the theory should not depend on the auxiliary `foliation'  of spacetime. As we will see in this section, there are multiple issues with this interpretation. Famously, \eqref{equ:Hamiltonian} also contains the generator of time evolution.    In other words, time evolution becomes inextricably mixed with a certain type of gauge-freedom, leading some to conclude that in GR evolution is ``pure gauge".  
This is one facet of what people have called ``the problem of time" (see e.g. \cite{Isham_POT, PoT_Anderson}).   It is related to the picture of ``block time" -- the notion that the object one deals with in general relativity is the \emph{entire} spacetime, for which the distinction between past, present and future is not fundamental. The worry is that  the Hamiltonian formalism  might be freezing the bathwater with the baby still inside. What can we salvage in terms of true evolution? 

 Even in the simplest example, it is not clear that the \lq\lq{}refoliation invariance\rq\rq{} interpretation is tenable, as shown by Torre et al \cite{Torre_evolution}. For a scalar field propagating in Minkowski spacetime  between two fixed hypersurfaces,    different choices of interpolating foliations will have unitarily inequivalent Schroedinger time evolutions. 
 
 But what about more broadly? What can we say about the attempt to represent relativity of simultaneity in the Hamiltonian --- both classical and quantum mechanical --- setting?  In this section I will investigate this question. The conclusion will be that, at least from the quantum mechanical perspective, it might make little sense to implement refoliation invariance.   I contend that this is because local time reparametrization represents an effective, but not fundamental symmetry.  I.e. I contend that \emph{invariance under refoliation  is not present at a quantum mechanical level, but should be recovered dynamically for states that are nearly classical.}
 
  In the following two subsection, I will expand first on obstacles for a quantization of refoliation symmetry in the Hamiltonian setting, and then in the Lagrangian setting.  I will expound on how these two types of obstacles point to timelessness as a possible resolution.  

\subsubsection{Hamiltonian evolution.}
Using the \lq{}covariant symplectic formalism\rq{}, one can  geometrically project symmetries of the Lagrangian theory onto the Hamiltonian framework \cite{Wald_Lee}. Using this formalism, one can precisely track how Lagrangian symmetries in the covariant field-space are represented as Hamiltonian flows in phase space. As shown by Wald and Lee \cite{Wald_Lee}, for this projection to be well-defined in the case of non-spatial diffeomorphism invariance (acting as a spacetime Lagrangian symmetry),  one needs to restrict the Lagrangian theory to consider only those fields which satisfy the equations of motion. Once restricted, indeed there exists a projection of the non-spatial diffeomorphism symmetry to the symplectic flow of the standard scalar ADM constraint \eqref{equ:Hamiltonian}. But otherwise, there isn\rq{}t such a correspondence. In other words,  the Hamiltonian formalism embodies the sacred principle of relativity of simultaneity only on-shell. 

In fact, because the ADM Hamiltonian is purely made of constraints, satisfying the constraints at every time and point in space means that the resulting data, $(g_{ab}(t, x), \pi ^{ab}(t, x)$ (and Lagrange multipliers), satisfy \textit{all} the equations of motion. In sum, imposing the Hamiltonian symmetries implies one must also impose (in Hamiltonian form) the Einstein equations of motion. Schematically, $H(x,t)=0\, \forall x,t\,\Rightarrow \dot \pi^{ab}=\{H, \pi_{ab}\}$ and  $\dot g_{ab}=\{H, g_{ab}\}$. Once the dynamics of the gravitational field are included in the Hamiltonian formalism  (home of canonical quantum mechanics), it is impossible to enforce relativity of simultaneity without also enforcing the gravitational equations of motion.

This is worth repeating: for generic off-shell spacetimes, the Hamiltonian constraint does not represent relativity of simultaneity. But symmetries should hold at the quantum level irrespectively of the classical equations of motion.  Which brings forth the question: what would it even mean to naively quantize \eqref{equ:Hamiltonian}?  What property would we be trying to represent with its quantization?

      If one nonetheless ignores these issues and pushes quantization, one gets the infamous Wheeler-DeWitt equation: 
\be \label{equ:wdw}
\hat H \psi[g]=0
\ee
where $\psi[g]$ is a wave-functional over the space of three-geometries.  Following this route, we see the classical `problem of time'  transported into the quantum regime:\footnote{ For the expert readers, I should note that a derivation of \eqref{equ:wdw} exists from the path integral formalism \cite{Hartle_Halliwell}. But that derivation already assumes that the Hamiltonian symmetries act on all the variables in the path integral, and thus it does not resolve the problem Wald and Lee pointed to.  Even ignoring all of these issues, equation \eqref{equ:wdw} has some further problems of its own. It has operator ordering ambiguities, functional derivatives of the metric acting at a singular point, no suitable inner product on the respective Hilbert space with respectable invariance properties, etc. 
 One could  also attempt to interpret \eqref{equ:wdw} as a Klein-Gordon equation with mass term proportional to the spatial Ricci scalar, but unlike Klein-Gordon, it is already supposed to be a quantized equation. Furthermore, the problem in defining suitable inner products are an obstacle in separating out a positive and negative spectrum of the Klein-Gordon operator \cite{PoT_Anderson}. \label{footnote:experts}} 
 One could look at equation \eqref{equ:wdw} as a \emph{time-independent}  Schr{\"o}dinger equation, which brings us again to the notion of  ``frozen time", from \eqref{equ:Schroedinger}. A solution of the equation will not be subject to time evolution; it will give a frozen probability wave-function on the space of three-geometries.
Other canonical approaches have so far similarly found insurmountable  problems  with the quantization of this constraint. 
I contend that it should simply not be quantized, as it does not represent a fundamental symmetry principle of the quantum theory.

 \subsubsection{Covariant quantum gravity}
 
 At a more formal level, to combine  \eqref{equ:GR} with our principle of superposition one should keep in mind that space-times define causal structures, and  it is far from clear how one should think about these in a state of superposition. For instance, which causal structure should one use in an algebraic quantum field theory approach when declaring that space-like separated operators commute? 
 Quantum field theory is formulated in a fixed spacetime geometry, while in general relativity spacetime is dynamical. Without a fixed definition of time or an a priori distinction between past and future, it is hard to impose causality or interpret probabilities in quantum mechanics.

    In quantum mechanics, we have time-evolution operators $e^{-i\hat H t}$, taking us from an initial physical state to a final one. In the language of path integrals, it is more convenient to express evolution in terms of a gauge-fixed propagator (the inversion of the quantum mechanical propagator requires gauge-degrees of freedom to have already been gauge-fixed)  $W(\phi_1, \phi_2)$, where $\phi$ (e.g. $\phi=(x,t)$) is  deemed to at least contain all the gauge-invariant information.\footnote{One can parametrize observables by families of gauge-fixings; as in e.g.: partial observables, in the sense of Rovelli \cite{Rovelli_partial}. Their inclusion does not change the discussion. } This simple characterization already raises two types of difficulties. 
 
The first is that there is no known local parametrization of physical (observable) Lorentzian 4-geometries, $[{}^\st{(4)}g]$ (global obstructions --- e.g. the Gribov problem \cite{Singer_Gribov} --- also exist, but are less concerning).  To understand what this means, we need to introduce the notion of a \lq{}slice\rq{}. A \lq{}slice\rq{} is a split between the physically equivalent (or gauge-equivalent) field configurations and the physically distinct ones. A local slice is one that performs this split only locally in field space, and it is equivalent to a local gauge-fixing if gauge-transformations form well-defined gauge-orbits (see figure 1).  Finding a slice theorem is relatively straightforward in the case of Riemannian metrics (Euclidean signature), for any dimension. 
 \begin{figure}\begin{center}
\includegraphics[width=.4\textwidth]{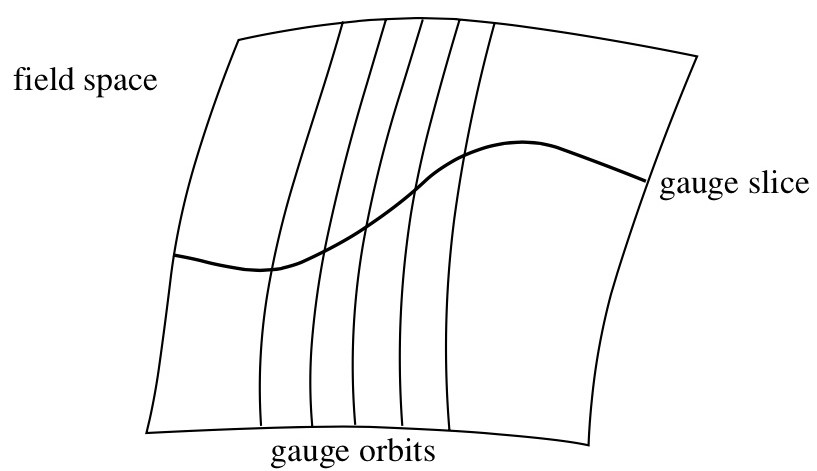}
\end{center}
\caption{A \lq{}gauge-slice\rq{} in the field-space of a field theory with gauge-symmetry.}
\end{figure}
To make a long story short, what goes wrong in the Lorentzian case involves the lack of invertibility of certain second order differential operators, which are elliptic in the Riemannian case.  In the case of Lorentzian metrics, the analogous operators are hyperbolic, invalidating the construction of the slice. It is the sign of Time that gets in the way.

  Indeed, a local parametrization, or slice for Lorentzian metrics has only been constructed  (by Isenberg and Marsden) for those Lorentzian metrics which i) satisfy the Einstein equations and ii) which furthermore admit a particular kind of time-foliation \cite{IsMa1982}.    This choice ---  Constant Mean Curvature (CMC) --- corresponds to synchronizing clocks  so that they measure the same expansion rate of space everywhere (i.e. same local Hubble parameter).

Of course, this is sub-optimal, to say the least. Quantization requires that we consider metrics which are off-shell, i.e. which do not obey the classical equations of motion. This issue  is  similar to the previously mentioned one, also surrounding relativity of simultaneity --- it only has a Hamiltonian representation for spacetimes satisfying Einstein equations.

  One source of problems is that, in GR, gauge-symmetries are spacetime diffeomorphisms, and the gravitational degrees of freedom that are (non-perturbatively) gauge-invariant thus can correspond to the entire spacetime, e.g. 
 $$\mbox{Diff-invariant quantities:} ~~~\int d^4x \sqrt{{}^\st{(4)}g} {}^\st{(4)}R,~~~  \int d^4x \sqrt{{}^\st{(4)}g}\,~ {}^\st{(4)}R^{\mu\nu}\,\, {}^\st{(4)}R_{\mu\nu}\,\,,~\mbox{etc}$$
Naively combining the two properties would require us to associate a quantum \lq\lq{}time-evolution\rq\rq{} between two different \textit{complete spacetimes}, not between observables within time.
For instance, the fundamental transition amplitude would be of the form $W([{}^\st{(4)}g_1], [{}^\st{(4)}g_2])$, where the square-brackets $[{}^\st{(4)}g]$ signify the quotient of the spacetime metric by all possible gauge-transformations (the spacetime geometry corresponding to the 4-metric ${}^\st{(4)}g$).  

To perhaps bluntly summarize the issue, it seems that non-locality in time requires prescience of what is to come. Although this non-locality could be compatible with a fully deterministic evolution, it is incompatible with the type of indeterminism inherent in quantum theory.  
 
\subsubsection{Gluing transition amplitudes for spacetime regions.} 

The previous argument assumes that spacetime does not possess boundaries, where the gauge-symmetries can be fixed. If one would like to have a piecewise approach --- that is, considering spacetime regions and then gluing them --- other types of questions arise about how to glue different transition amplitudes in gauge-theories, for instance, questions about the compatibility of  gauge-fixing the degrees of freedom at the boundaries \cite{Donnelly_2016, Donnelly_local} (which is problematic for a variety of reasons, including the previous one of a lack of slice). 

In a 3+1 decomposition, at least two related difficulties arise for the path integral. One could take $\dot N=0=N_i$ , for lapse $N$ and shift $N_i$. This is part of the standard gauge-fixing taken for transition amplitudes in GR, such as for Hartle-Hawking \cite{Teitelboim_path}. But these won't cover generic spacetimes; such coordinates generically form caustics. This is related to the problem mentioned above: there is no (known) slice theorem for Lorentzian metrics \cite{IsMa1982}.  

Moreover, as mentioned before (in footnote \ref{footnote:experts}), the proof that such transition amplitudes obey the Hamiltonian invariance equations, \eqref{equ:wdw},  assumes invariance under the actions of the constraints, i.e. it assumes phase space invariance, not spacetime (Lagrangian) invariance. Thus the argument expounded above with regards to the Wheeler-DeWitt equation applies also here: it is unclear if the transition amplitude expresses relativity of simultaneity \cite{Wald_Lee}. 

The second difficulty is that the spacetime corresponding to each gauge-fixed path is taken to have boundaries (at least the initial and final time-slices), and, again,  the fate of diffeomorphisms in the presence of boundaries is a very current matter of discussion in the community.\footnote{These concerns have resurfaced due to the study of entanglement entropy \cite{Donnelly_entanglement}.} At least in the presence of degrees of freedom which are not strictly topological in nature, that is.  

Indeed, many approaches to quantum gravity, such as spin foams (see \cite{Rovelli_book} and references therein) are inspired by an approach to topological quantum field theory (TQFT) \, originated by Atyiah and Segal \cite{Atiyah1988}. These approaches depict a transition amplitude from (co)boundaries of a manifold. The boundaries host  states, and the interior of the manifold encodes the transition amplitude between these states. The transition amplitude can be obtained by a path integral of all field histories between the two boundaries. And it is true that there are many examples for which we understand quantization of TQFT\rq{}s with diffeomorphism symmetry, including gravity in 2+1 spacetime dimensions (see \cite{Carlip_liv_rev}). However, being topological in nature, in none of these theories is there a discrepancy between the full field space and that subset which satisfies the equations of motion; the equations of motion are trivially satisfied by using gauge transformations; in this case indeed there is no difference between implementing symmetries on-shell or off-shell. Therefore, in these theories, the counter-arguments outlined in this section are not valid; it requires a gap between what is kinematical and what is dynamical. 

\subsubsection{Many problems...}
Let us take stock of the many problems related to time in GR which we have mentioned: first, `dynamics' ---  the Einstein equations  -- has little to do with causality properties; even kinematically, the field $g_{\mu\nu}$ already carries causal relations.\footnote{As a background, that is. One can move to the 3+1 context,  in which, after gauge-fixing, one finds  hyperbolic equations of motion propagating field disturbances. But then, one is back to the other issues I have mentioned: first, regarding generic gauge-fixings of the 4-diffeomorphisms. Secondly, regarding the inextricable relation between evolution and gauge symmetry; is it possible to satisfy one but not the other?} Second, the Hamiltonian only relates to redefinitions of simultaneity when the Einstein field equations are met \cite{Wald_Lee}. Relatedly, there is no middle ground: if  the Hamiltonian constraint is satisfied everywhere then  the full equations of motion are also satisfied; the lines between symmetry and evolution are completely blurred. In other words, within the Hamiltonian formalism -- the most natural formalism for quantum mechanics --  one cannot implement the symmetries off-shell, that is, in a truly quantum mechanical manner.  Indeed, as mentioned in the beginning of the section, even the simplest example field-theory (with local degrees of freedom) testing  invariance under refoliations  in the quantum mechanical realm is problematic  \cite{Torre_evolution}. Thirdly, in the covariant approach, no generic (i.e. also off-shell) gauge-fixing is known. This is again a problem of the signature of operators related to the time direction.

\subsection{Symmetries, relationalism and `laws of the instant'}\label{sec:symmetries}

   Many of the problems in the previous section seem to point in the same direction: parametrizing  physical degrees of freedom in the presence of relativity of simultaneity is a difficult task.  Perhaps restricting the fundamental symmetries of the theory to disallow this mixing with evolution would cure some of these issues. As I have emphasized, we only need to recover refoliation invariance  on-shell, i.e. only after the equations of motion have been imposed. This frees up the theory to accept different fundamental symmetry principles, and delegate the fulfillment of  refoliation invariance to on-shell properties. In this section, we will uncover what this new sort of symmetries can be. 
  
  \subsubsection{A silver lining.}
 There are makeshift patches to the problems of time we have mentioned, and in them,  we can find a silver lining. For instance, as I mentioned,  a very weak form of a slice theorem --- essentially generic gauge-fixings--- exists for GR \cite{IsMa1982}. It requires that the spacetime satisfy the Einstein equations and admit a constant-mean-curvature (CMC) foliation, i.e. synchronizing clocks so that the expansion of space is constant everywhere. Moreover, going to the 3+1 framework to study \textit{dynamics}, one finds that {most  formal proofs for existence and uniqueness of solutions  also require the use of CMC foliations}, through the so-called York method  \cite{York, York2}. Indeed, most if not all of the numerical simulations used to model black hole mergers have worked within CMC \cite{Pretorius_binary}, even helping to interpret LIGO data; and every test that has ever been passed by GR is known to be consistent with a CMC foliation. 

  The constraints these foliations need to satisfy generate certain dynamical (symplectic) flows: changes of spatial scale, i.e.  \textit{local conformal transformations}, as we will see shortly.
Indeed,  CMC foliations  have special properties in GR:  the evolution of the spatial conformal geometry decouples from the evolution of the pure scale degrees of freedom of the metric. 
     Surprisingly,  both the York method and the slice theorem show that \textit{although GR is not fundamentally concerned with spatial conformal  geometries, it is deeply related to them}. As we now comment on, this is not an accident; the most general sort of symmetries that act pointwise in configuration space and locally in space are indeed conformal transformations and diffeomorphisms.  

In the following  subsection, I introduce these symmetries. In subsection \ref{sec:laws}, I describe this result: these symmetries are the most general ones with an inherently \lq{}instantaneous\rq{}, local, action.
\subsubsection{Relationalism}
In Hamiltonian language, the most general symmetry transformation acts through the Poisson bracket $\{\,\cdot\,\,\cdot\,\}$, on configurations 
 \be \delta_\epsilon g_{ij}(x)=\left\{\int d^3 x\, F[g,\pi;x\rq{})\epsilon(x\rq{})\,,\,g_{ij}(x)\right\}
 \ee where $\epsilon$ is the not-necessarily scalar gauge parameter, which in this infinite dimensional context is a function on the closed spatial manifold, $M$, and we are using DeWitt\rq{}s mixed functional dependence, i.e. $F$ depends functionally on $g_{ij}$ (not just on its value at $x\rq{}$), as denoted by square brackets,  but it yields a function with position dependence -- the \lq\lq{} $;x\rq{})$\rq\rq{}  at the end.
 
Regarding the presence of gauge symmetries in configuration space, we would like to implement the most general relational principles that are applicable to space (as opposed to spacetime). At face value, the strictly relational symmetries should  be: 
\begin{itemize}
\item {\bf Relationalism of locations.} In Newtonian particle mechanics this would imply that only relative positions and velocities of the particles, not their absolute position and motion, are relevant for dynamics. In the center of mass frame, i.e. in which the total linear momentum vanishes ($\vec P=0$), the supposition only holds if the total angular momentum of the system also vanishes (see \cite{Barbour_Mach, Flavio_tutorial}), $\vec L\approx 0$. In the gravitational  field theory case, relationalism of locations  is represented by the (spatial) diffeomorphism group  Diff$(M)$ of the manifold $M$. It is generated by a constraint $F[g,\pi;x\rq{})=\nabla_i\pi^i_{~j}(x)\approx 0$, which yields on configuration space the transformation $\delta_{\vec\epsilon} g_{ij}(x)= \mathcal{L}_{\vec\epsilon}g_{ij}(x)$, which is just the infinitesimal dragging of tensors by a diffeomorphism. 

\item {\bf Relationalism of scale.} In Newtonian particle mechanics this relational symmetry would imply that  only the relative distance of the particles, not the absolute scale, is relevant for dynamics. It only holds if the total \lq{}dilatational momentum\rq{} of the system vanishes (see \cite{Barbour_Mach, Flavio_tutorial}). In the gravitational field theory this symmetry is represented by the group of scale transformations (also called the Weyl group), $\mathcal{C}(M)$, which is symplectically generated by $F[g,\pi;x\rq{})=g_{ij}\pi^{ij}(x)\approx 0$ (it yields on configuration space the scale transformation $\delta_\epsilon g_{ij}(x)=\epsilon(x)g_{ij}(x)$). In this case the infinitesimal gauge parameter $\epsilon$ is a scalar function, as opposed to a vector field $\vec{\epsilon}$ for the diffeomorphisms.  

\end{itemize}
Unlike what is the case with the constraints emerging from the Hamiltonian ADM formalism of general relativity, these symmetries form a (infinite-dimensional)  closed Lie algebra.  

\subsubsection{Laws of the instant}\label{sec:laws}
 Even if we disregard considerations about relationalism,  local time-reparametrizations, or refoliations, also don't act as a group in spatial configuration space, and thus do not allow one to form a gauge-invariant  quotient from its action. That is, given a particular linear combination -- defined by a smearing\footnote{The $\perp$ notation is standard to represent parameter acting transversally to the constant-time surfaces.} $\lambda^\perp$ -- of the Hamiltonian constraints given in \eqref{equ:Hamiltonian}, it generates the following transformation:
\be\label{refol_sym}\delta_{\lambda^\perp}g_{ab}(x)=\frac{2\lambda^\perp (\pi_{ab}-\frac12 \pi g_{ab} )  }{\sqrt{g}}(x)\ee which depends not only on the metric, but also on the momenta. This dependence is clearly in contrast to the symmetries related to  relationalism of scale and of position, above. It means that the 3-metric by itself carries no gauge-invariant information. 

 Indeed, for  the associated symmetry to have an action on configuration space that is independent of the momenta, a given constraint $F[g,\pi,\lambda]\approx 0$ must be linear in the momenta. This already severely restricts the forms of the functional to:\footnote{Up to canonical transformations which don't change the metric, i.e. with generating functionals of the form $\int d^3 x\,  (\tilde\pi^{ab}g_{ab}+\sqrt{g}F[g])$, for $F$ any functional of $g$ and $\tilde \pi^{ab}$ the new momentum variable. } 
\be\label{equ:total_symmetry}F[g,\pi,\lambda]=\int \tilde F(g,\lambda)_{ab}(x)\pi^{ab}(x)\ee
so that the infinitesimal gauge transformation for the gauge-parameter $\lambda$ gives:
$$\delta_\lambda g_{ab}(x)=\tilde{F}(g,\lambda)_{ab}(x).$$

Thus I would like symmetries to act solely on configuration space. Only such symmetries are compatible with the demand that the $W(g^1_{ab}\,,g^2_{ab})$ give all the information we need about a theory, since only such symmetries allow $g_{ab}$ to carry gauge-invariant information.   I thus require that  $\delta_\epsilon g^1_{ab}(x)=G[g^1_{ab}\,,\epsilon\,;x)$ for some mixed functional $G$ -- which crucially \emph{only} depends on $g_{ab}^1$. 
 
   In other words, the action of the symmetry transformations of \lq{}now\rq{}, only  depend on the content of \lq{}now\rq{}. 
The action of these relational symmetries on each configuration $g_{ab}$ is self-determined,  they do not depend on the history of the configuration or on configurations $\tilde g_{ab}\neq g_{ab}$. In  appendix \ref{sec:LOTI}, I sketch a proof  that  the relational symmetries of scale and position  are indeed the only symmetries whose action in phase space projects down to an intrinsic action on configuration space.

The conclusion of this argument is that spatial relationalism is singled out by demanding that symmetries have an intrinsic action on configuration space. Just to be clear, this feature is not realized by the action of the ADM scalar constraint \eqref{equ:Hamiltonian}, since it is a symmetry generated by terms quadratic in the momenta and thus the transformation it generates on the metric requires knowledge of the conjugate momentum (and vice-versa).

 Lastly (and also unlike what is the case with the scalar constraint \eqref{equ:Hamiltonian}),\footnote{Barring the occurrence of metrics with non-trivial isometry group.}  the action of these symmetries endows configuration space $\mathcal{M}$ with a well-defined, neat principal fiber bundle structure  (see \cite{FiMa77}), which enables their quantum treatment \cite{Conformal_geodesic}.

 \section{Timelessness, quantum mechanics, and configuration space} \label{canonical_timeless}
 
   As mentioned, any theory that is completely relational would preferably not contain an explicit, Newtonian time variable. Briefly summarized, relationalism is the belief  that all relevant physical information, including Time, should be deducible from the relation between physical objects. Thus the presence of an \lq\lq{}external time\rq\rq{} is odd from the point of view of relationalism, and should be extractable from internal properties of curves in configuration space. We thus turn now to timelessness.

What does a timeless, relational theory, quantum or classical, look like? A long literature exists on this matter, and it is of course beyond the scope of this work to give any reasonably detailed account of the subject. 

Instead, I will start with a brief description of  pre-scientific debates surrounding the issue, in section \ref{sec:Topology}.
Skipping straight into modern ideas about timeless quantum mechanics, in section \ref{sec:Hamiltonian_timeless} I will then give a very brief account of the results of \cite{Chiou} which are specially useful for my purposes. Chiou translates canonical timeless quantum mechanics (see e.g. \cite{Rovelli_book}, briefly summarized in appendix \ref{sec:canonical_timeless}) into  the path integral formulation --  which is the approach I believe carries the most useful conceptual baggage. I will end the section by illustrating the main features of  a path integral for quantum gravity satisfying such principles. 

\subsection{The special existence of the present - Parmenides and Zeno}\label{sec:Parmenides}\label{sec:Topology}

It could be argued that we do not ``experience"  space-times. We experience `one instant at a time', so to say. We of course still appear to experience the passage of time, or perhaps more accurately, we (indirectly) experience changes in the spatial configuration of the world around us, through changes of the spatial configuration of our brain states. 

But if present experience is somehow distinguished, how does ``change" come about?  This is where Parmenides has something to say that is relevant for our discussion. Parmenides was part of a group called the Eleatics, whose most prominent members were himself and Zeno, and whose central belief was that all change is illusory. 
The reasoning that led them to this conclusion was the following: if the future (or past) is real, and the future is not existing now, it would have both properties of existing and not existing, a contradiction (or a `turning back on itself').  Without past and future, the past cannot transmute itself into future, and thus there is also no possible change. Of course, the argument hinges on the distinction we perceive between present, past and future.


Many centuries later, Augustine of Hippo  picked up the question, concluding that change was an illusion and yet,
\begin{quote}
How can the past and future be, when the past no longer is, and the future is not yet? As for the present, if it were always present and never moved on to become the past, it would not be time, but eternity.[...] 
Nevertheless we do measure time. We cannot measure it if it is not yet into being, or if it is no longer in being, or if it has no duration, or if it has no beginning and no end. Therefore we measure neither the future nor the past nor the present nor time that is passing. Yet we do measure time.\end{quote} 
According to Augustine, Time is a human invention: the difference between  future and past is merely the one between anticipation and memory. 

To the extent that future and past events are real, they are real now, i.e. they are somehow encoded in the present configuration of the Universe. Apart from that, they can be argued not to exist. My memory of the donut I had for breakfast is etched into patterns of electric and chemical configurations of my brain, \emph{right now}. We infer the past existence of dinosaurs because it is encoded in the genes of present species and in fossils in the soil. In a timeless Universe, what we actually do is deduce from the present that there exists a continuous curve of configurations connecting \lq{}now\rq{} to some other configuration we call `the past'.  But if the ideas of past and future were all completely immaterial, how would we have come to have such illusions in the first place? They surely are not false, and other instants should exist. But then how to connect a  snapshot of the dinosaur dying with the snapshot of the archaeologist finding its remains?  Does the Eleatic argument bring about a `solipsism of the instant'?

\subsubsection{A more mathematical posing of the question.}
 Let's call configuration space $\mathcal{M}$, which we endow with a reasonable topology. Given an appropriate action functional over $\mathcal{M}$, one obtains continuous curves that extremize this functional. It makes sense to have configuration `bite on donut'   connected by one such continuous curve to configuration `me, reminiscing about donut, six hours later'. But what is the meaning of these curves? In which sense can we think of ourselves as traversing them?

We now turn to the meaning of  timeless configuration space,  how to construct a theory of quantum mechanics there, and how a notion of traversing such a curve arises.

\subsection{Timeless path integral in quantum mechanics}\label{sec:Hamiltonian_timeless}

We start with a finite-dimensional system, whose configuration space, $\mathcal{M}$, is coordinitized by $q^a$, for $a=1,\cdots, n$. An observation yields a complete set of $q^a$, which is called an event.  Let us start by making it clear that no coordinate, or function of coordinates, need single itself out as a reference parameter of curves in $\mathcal{M}$. The systems we are considering are not necessarily `deparametrizable' -- they do not necessarily possess a suitable notion of time variable. 

Now let $\Omega=T^*\mathcal{M}$ be the cotangent bundle to configuration space, with coordinates $q^a$ and their momenta $p_a$. The classical dynamics of a reparametrization invariant system is fully determined once one fixes the Hamiltonian constraint surface in $\Omega$, given by $H=0$.
A curve $\gamma\in \mathcal{M}$ is a classical history connecting the events $q^a_1$ and $q^a_2$ if there exists an \emph{unparametrized} curve $\bar\gamma$ in $T^*\mathcal{M}$ such that the following action is extremized:
\be S[\bar\gamma]=\int_{\bar\gamma}p_adq^a 
\ee
for curves lying on the constraint surface $H(q^a, p_a)=0$, and are such that $\bar\gamma$'s projection to $\mathcal{M}$ is $\gamma$, connecting $q^a_1$ and $q^a_2$. 

 Feynman's original demonstration of the equivalence between the standard form of non-relativistic quantum mechanics and his own path integral formulation relied on refining time slicings. The availability of time  gave a straightforward manner by which to partition paths into smaller and smaller segments. Without absolute time, one must employ new tools in seeking to show the equivalence. For instance, a parametrized curve $\bar\gamma:[0,1]\rightarrow \Omega$ need not be injective on its image (it may go back and forth). This requires one to use a Riemann-Stieltjes integral as opposed to a Riemann one in order to make sense of the limiting procedure to infinite sub-divisions of the parametrization. In the end, a timeless transition amplitude  becomes  (see appendix \ref{sec:canonical_timeless}):
\be
\label{equ:path_integral_phase}
W(q_1,q_2)=\int \mathcal{D}q^a\int \mathcal{D}p_a\, \delta[H] \exp{\left[\frac{i}{\hbar}\int_{\bar\gamma}p_adq^ a\right]}
\ee
where the path integral sums over paths whose projection starts at $q_1$ and ends at $q_2$, and $H$ is a single reparametrization constraint. In the presence of gauge symmetries, if it is the case that these symmetries form a closed Lie algebra, one can in principle use the group averaging procedure mentioned in appendix \ref{sec:canonical_timeless}, provided one uses a similarly translation invariant measure of integration (which is available for the single, or global, reparametrization group).

For a strictly deparametrizable system,\footnote{I.e. one for which $[\hat H(t_1), \hat{H}(t_2)]=0$.  If this is not the case, the equality will only hold semi-classically.} one obtains again: 
$$ W(t_1,q_1^i,t_2,q^i_2)\sim \int \mathcal{D} t\,  G(t_1,q_1^i,t_2,q^i_2) \sim G(t_1,q_1^i,t_2,q^i_2)
$$ up to an irrelevant overall factor. Further, if the Hamiltonian is quadratic in the momenta, one can integrate them out and obtain the configuration space path integral with the Lagrangian form of the action.

\subsubsection{The absence of change.}
For a theory that contains some driver of change, an absolute Time of some sort, we would extend our configuration space with an independent time variable, $t$, making the system effectively deparametrizable. With this absolute notion of Time, and ontological deparametrization of the system, evolution from $t_1$ to $t_2$ would not require any further definition. At this point we could stop, claiming that we have expounded on what we expect a relational theory of space to look like. We would be able to define a Schrodinger equation as in the usual time-dependent framework, and go about our business.  Shape dynamics employing the complete relational symmetries is a theory of that sort.\footnote{The absolute time used in the original version of shape dynamics \cite{SD_first}, is of the form $\langle\pi^{ab}g_{ab}\rangle$, where brackets denote the spatial average. This quantity is only invariant wrt Weyl transformations that preserve the total volume of space, and is thus not completely relational. One can extend the conformal transformation to the full group, acquiring an absolute time parametrization \cite{Tim_duration}. } However, the presence of Time there is still  disturbing from a relational point of view: where is this Time if not in the relations between elements of the configurations? Therefore, to fully satisfy our relational fetishes, we must again tackle the question posed at the introduction: \textit{without a driver for change, what is the meaning of  a transition amplitude?} The answer will come in section \ref{sec:records}, where we introduce \lq{}records\rq{}. Before that, we need to transpose the mechanisms of this section to the quantum gravity context.



\subsubsection{A relational transition amplitude for Quantum Gravity.}\label{sec:timeless_QG}

In this section, I reviewed a timeless path integral formulation of quantum mechanics by Chiou \cite{Chiou}, building on previous results on timeless quantum mechanics (see \cite{Rovelli_book} for a review). In these formulations,  configuration space is the `space of all possible instants'.  

Configuration space for timeless field theories, which I will still denote by $\mathcal{M}$,  is the set of all possible field configurations over a given closed manifold $M$.  Each point of configuration space $q\in \mathcal{M}$ is a ``snapshot" of the whole Universe.\footnote{ For instance, it could be the space of sections on a tensor bundle, $\mathcal{M}=C^\infty(TM\otimes\cdots TM\otimes TM^*\cdots TM^*)$. In the case of gravity, these are sections of the positive symmetric tensor bundle: $\mathcal{M}=C_+^\infty(TM^*\otimes_S TM^*)$.} I will require symmetries to be \lq{}laws of the instant\rq{} precisely so that they are compatible with a theory  defined at its most fundamental level by $W(q_1,q_2)$.  

Given these symmetries (and the principal fiber bundle structure they form), we take the analogous of \eqref{equ:path_integral_phase}, schematically projected down onto the space of conformal geometries\footnote{The full treatment of the gauge conditions requires a gauge-fixed BRST formalism, which is a level of detail I don't need here.  See \cite{Conformal_geodesic} for a more precise definition, equation (28), where we use $K(g_1, [g_2])$ as opposed to $W([g_1],[g_2])$. }:
\be
\label{equ:path_integral_conformal_geometrodynamics}
W([g_1],[g_2])=\int \mathcal{D}[g]\int \mathcal{D}[\pi]\,  \exp{\left[\frac{i}{\hbar}\int_{\bar\gamma}[\pi^{ab}]\,\delta {[g_{ab}]}\right]\delta H([g],[\pi])}
\ee
where I have (again, schematically) used square brackets to denote the conformal-diffeo equivalence classes of the metric and momenta, and where $\delta H$ represents a single reparametrization constraint (not an infinite amount, as in the ADM scalar constraint). Schematically, this transition amplitude should play the role of \eqref{equ:path_integral_phase} in the field theory case. 

Although considerations of quantum gravity and its problems have led us to value the timeless, or instantaneous, representation of quantum theory, having reached this point, we revert back to the more general notation,   denoting the equivalence classes $[g]$ and all other equivalence classes of fields under the appropriate instantaneous symmetries,  by the standard coordinate variable, $q$.


\section{Records and timelessness}\label{sec:records}
Suppose then, that we have in our hands a $W(q_q, q_2)$ for which $q$'s carry also the gauge-invariant degrees of freedom. 
Still, as stressed in previous sections; without some \lq{}driver of change\rq{} --- which we usually call time --- what is the meaning of this transition amplitude? 
 Here we will see how such a meaning can arise from timelessness. The ultimate meaning $W(q_1,q_2)$ can give rise to is simple:  the likelihood that records of $q_1$ will be found in $q_2$.  In section \ref{sec:Born}, I will build the scaffolding for a static volume-form in configuration space. This requires a definition of the space of \lq{}beables\rq{} and of an \lq{}anchor\rq{}  to the path integral. Having done this, in section \ref{sec:records} I introduce the structure which allows one to ascribe histories to properties of the static volume-form -- -records. However, records are not enough to talk about conservation of probability, and at the end of the section I sketch how this can be done. 

\subsection{Born rule and the preferred configuration.}\label{sec:Born}

\subsubsection{The ontological status of configuration space.}
In a true spatially relational theory, an  instantaneous state of an observer is encoded in a partial field configuration.   There are no subjective overtones  attributed to an observer -- it is merely a (partial) state of the fields. Of course, there are many regions of configuration space where no such thing as an observer will be represented. 

Since each point is a possible `now', and there is no evolution, each `now'  has an equal  claim on existing. This establishes the plane of existence, every `now\rq{} that can exist, does exist!  We are at least partway towards the adage of quantum mechanics. If this was a discrete space, we could say that each element has the same weight. This is known as the principle of indifference and it implies that we count each copy of a similar observer once.\footnote{The intuition obtained for Many Worlds in the discrete configuration spaces can be misleading for our purposes. In that case, each `branch\rq{} can be counted, and one needs a further explanation to count them according to the Born rule. There are different ways of going about this, e.g.: based on this principle, and on the Epistemic Principle of Separability, Carroll et al claim that the Born rule can be derived \cite{Carroll}.  } 

But configuration space is a continuous space, like $\mathbb{R}^2$ (but infinite-dimensional). Unlike what is the case with discrete spaces, there is  no preferred way of counting points of $\mathbb{R}^2$. We need to imprint $\mathcal{M}$ with a volume form; each volume form represents a different way of counting configurations. 

\subsubsection{Born rule.}
Contrary to what occurs in standard time-dependent Many Worlds quantum mechanics, I will define a single, standard time-independent `volume element' over configuration space $\mathcal{M}$.  Integrated over a given region, this volume element will simply give the volume, or the amount, of configurations in that region.\footnote{Of course, these volume forms are divergent and  technically difficult to define. Properties of locality of the volume form, discussed in \cite{Locality_riem} are essential to show that nonetheless their definition reduces to the usual Born rule for isolated finite-dimensional systems.  Furthermore, only ratios of the volume form have any meaning. } 

The volume form $P(q)Dq$ is defined as a positive scalar function of the transition amplitude, $P(q):=F(W(q^*,q))$. We need to explain the notation in this equation. First, there is still a sign of the principle of indifference in the manner we choose the \lq{}bare\rq{}  volume element, $Dq$; it is chosen as the translationally invariant measure in the field-theory context. But since it is being multiplied by  $F:\mathbb{C}\rightarrow \mathbb{R}^+$, it can still be anything. Now we endow $F$ with the extra  property 
 that it preserves the multiplicative group structure, 
 \be\label{equ:factorization_density} F(z_1z_2)=F(z_1)F(z_2)\ee 
an important property to recover locality and an empiric notion of records from the transition amplitude \cite{Locality_riem}. This measure, $F$, gives a way to ``count" configurations, and it is assumed to act as a  positive functional of the only non-trivial function we have defined on $\cal{M}$, namely, the transition amplitude $W(q^*,q)$. Together with certain locality properties of $W(q^*,q)$ discussed in \cite{Locality_riem},  equation \eqref{equ:factorization_density} together with the classical limit uniquely leads to a derivation of the \lq{}Born volume\rq{}:
 $F(W(q^*,q))=|W(q^*,q)|^2$, i.e.
\be\label{equ:volume_form}
P(q)=|W(q^*,q)|^2
\ee 
Lastly, in the definition of $P(q)$ I have sneaked in a `in'  configuration, $q^*$, which defines once and for all the static volume form over (reduced) configuration space. I define $q^*$ roughly as the simplest, most structureless configuration of the fields in question. Note that this can only be a meaningful statement if $q$ carries its own physical content; i.e. for symmetries which are laws of the instant. 

\subsubsection{The preferred configuration, $q^*$.}

This section gives a set of natural choices for $q^*$, depending on the configuration space, gauge group, and manifold topology.  Reduced configuration spaces may not form smooth manifolds, but only what are called stratified manifolds. This is because the symmetry group $\mathcal G$ in question -- whose action forms the equivalence relation by which we are quotienting -- may act \emph{qualitatively} differently on different orbits. If there are subgroups of the symmetry group  ---stabilizer subgroups--- whose action leave a point $\tilde q$ fixed, the symmetry does not act \lq\lq{}fully\rq\rq{} on $\tilde q$ (or on any other representative $\bar q\in [\tilde q]$); some subgroups of $\mathcal{G}$ simply fail to do anything to $\tilde q$. This implies  the quotient of configuration space wrt to the full symmetry may vary in dimensionality. 

  Taking the quotient by such wavering actions of the symmetry group creates a patchwork of manifolds. Each patch is called a stratum and  is indexed by the stabilizer subgroup of the symmetry group in question (e.g. isometries as a subgroup of Diff(M)). The larger the stabilizer group, the more it fails to act on $\tilde q$, the lower the dimensionality of the corresponding stratum.  The union of these patches, or strata,  is called a stratified manifold. It is a space that has nested \lq\lq{}corners\rq\rq{} -- each stratum has as boundaries a lesser dimensional stratum.  A useful picture to have in mind for this structure is  a cube (seen as a manifold with boundaries). The interior of the cube has boundaries which decomposes into faces, whose boundaries decompose into lines, whose boundaries decompose into points.  The higher the dimension of the boundary component, the smaller the isometry group that its constituents have,\footnote{E.g. let $\mathcal{M}_o$ be the set of metrics without isometries. This is a dense and open subset of $\mathcal{M}$, the space of smooth metrics over $M$. Let  $I_{n}$ be the isometry group of the metrics $g_n$, such that the dimension of $I_n$ is $d_n$. Then the quotient space of metrics with isometry group $I_n$ forms a manifold with boundaries, $\mathcal{M}_n/\mbox{Diff}(M)=\mathcal{S}_n$. The boundary of $\mathcal{S}_n$ decomposes into the union of $\mathcal{S}_{n\rq{}}$ for $n\rq{}>n$ (see \cite{Fischer}).}, i.e. the more fully $\mathcal{G}$ acts on it. Thus the interior of the cube would have no stabilizer subgroups associated to it, and the one-dimensional corners the highest dimensional stabilizer subgroups. Everything in between would follow this order:  the face of the cube could be associated to a lower dimensional stabilizer subgroup than the edges, and the edges a lower one than the corners. 

Configurations with the highest possible dimension of the stabilizer subgroup are what I define as $q^*$ -- they are the pointiest corners of this concatenated sequence of manifolds. And it is these topologically preferred singular points of configuration space that we define as an origin of the transition amplitude. Their simplicity coincides with ---or is a reinterpretation of--- characteristics we would expect from a low entropy beginning of the Universe. 

    Thus, depending on the symmetries acting of configuration space, and on the topology of $M$, one can have different such preferred configurations. 
    For the case at hand -- in which we have both scale and diffeomorphism symmetry and  $M=S^3$-- there exists two sorts of such preferred points, one connected to the rest of the quotient space and the other disconnected.  The preferred $q^*$ of $\mathcal{M}/($Diff$(M)\,\ltimes\mathcal{C})$ which is connected to the rest of the manifold is the one corresponding to  the round sphere. The disconnected point is the completely singular metric, $q^*=g_{ab}=0$.\footnote{For conformal transformation, we can see it is disconnected in the quotient space, because we have no access to $g_{ab}=0$. Choosing any given reference $\bar g_{ab}$, we can conformally project $g_{ab}$, i.e. $[g_{ab}]=\left(\frac{\bar g}{g}\right)^{1/3} g_{ab}$. As $g_{ab}$ becomes degenerate, its determinant goes to zero, and any such conformal projection diverges. Any such $[g_{ab}]$ is therefore, at bes, \lq{}infinitely far\rq{}.   }
    
    If we look at just the spatial spatial diffeomorphisms, then the natural choice becomes the singular metric $q^*=g_{ab}=0$. In the Hartle-Hawking state, in minisuperspace (where refoliations act as a single reparametrization, as they would here), this is (equivalent to) the initial state chosen.

\subsection{Records}\label{sec:records} 

\subsubsection{Semi-classical records.}
We are now in position to relinquish ``a driver of change''. With the notion of `records', about to be introduced, we can recover all the appearances of change, without it having to be introduced by fiat as extraneous structure. 

Having defined $q^*$, we can set it as $q_1$ and obtain a meaningful transition amplitude $W(q_1,q_2)$ to `now\rq{}, represented by $q_2$. At a fundamental level, $q^*$, together with a definition of $F$ and the action,  completely specify the physical content of the theory by giving the volume of configurations in a given region of $\mathcal{M}$.

  It is this anchoring of the amplitude on $q^*$ that allows probabilities to depend only on the \lq{}past\rq{}.  
   It is also what permits  the existence of another class of object which I call  \emph{records}. 

The system one should have in mind as an example of such a structure is  the Mott bubble chamber \cite{Mott}. In it, emitted particles from  $\alpha$-decay in a cloud chamber condense bubbles along their trajectories. A quantum mechanical treatment involving a timeless Schr{\"o}dinger equation finds that the wave-function peaks on configurations for which bubbles are formed collinearly with the source of the $\alpha$-decay. In this analogy, a `record holding configuration' would be any configuration with $n$ collinear condensed bubbles, and any configuration with $n'\leq n$ condensed bubbles along the same direction would be the respective  `record configuration'. In other words, the $n+1$-collinear bubbles configuration holds a record of the $n$-bubbles one.  For example, to leading order, the probability amplitude for  $n$ bubbles along the $\theta$ direction obeys 
\be\label{equ:Mott}P[(n,\theta),\cdots, (1,\theta)]\simeq P[(n\rq{},\theta),\cdots, (1,\theta)] P[(n\rq{},\theta),\cdots, (1,\theta)|(n,\theta),\cdots, (1,\theta)]\ee
 where $n\rq{}<n$, and $ P[A|B]$ is the conditional probability for $B$ given $A$. 

Let us sketch how this comes about in the present context. When semi-classical approximations may be made for the transition amplitude between $q^*$ and a given configuration, we have  
\be\label{equ:semi_classical}W_{\mbox{\tiny cl}}(q^*,q)=\sum_{\gamma_j}\Delta^{\frac{1}{2}}_j\exp{((i/\hbar)S_{\mbox{\tiny cl}}[\gamma_j])}\ee where the $\gamma_j$ are curves that extremize the action, which on-shell we wrote as $S_{\mbox{\tiny cl}}[\gamma_j])$,  and $\Delta$ are certain weights for each one (called Van-Vleck determinants\footnote{The weights of each extremal path are given by the Van-Vleck determinant, $\Delta_i=\frac{\delta \pi^i_1}{\delta \phi}$, where $\pi^i_1$ is the initial momentum required to reach that final $\phi$. Having small Van-Vleck determinant means that slight variations of the initial momentum give rise to large deviations in the final position. Let me illustrate the meaning  of a Van-Vleck determinant with a well-known heuristic example: suppose that $\phi_1$ contains a broken egg.  If $\phi$ represents a configuration with that same egg\footnote{Same for all practical purposes -- its relations to the rest of the configuration consist of what we would identify as the same egg.} unbroken (still connected to $\phi_1$ by an extremal curve), small deviations in initial velocity of configuration change at $\phi_1$ will result in a a final configuration very much different (very far from) $\phi$. }). This formula is approximately valid when the $S_{\mbox{\tiny cl}}[\gamma_j])>>\hbar$.

Roughly speaking, when all of $\gamma_j$ go through a configuration $q_r\neq q$, I will define $q$ as \emph{possessing a semi-classical record} of $q_r$. Note that this is a statement about $q$, i.e. it is $q$ that contains the record (a more precise definition is left for \cite{QG_deco}). I will call $\mathcal{M}_{(r)}$ the entire set that contains $q_r$ as a record.

For $q\in \mathcal{M}_{(r)}$, it can be shown that the amplitude suffers a decomposition (this is shown in  \cite{QG_deco})
 \be\label{equ:record_dec} W(q^*,q)\simeq W(q^*,q_r)W(q_r,q)
\ee
To show this in a simplified setting of a deparametrizable system --- i.e. when the Hamiltonian admits a split $H(q,p)=p_o+H_o(q_i, p^i, q_o)$, with $[H_o(t), H_o(t')]=0$ ---, one uses the same techniques as those used to prove the semi-group properties of the semi-classical amplitude. For $q_o=t$: 
\be W_{\mbox{\tiny cl}}((q^i_1,t_1),(q^i_3,t_3))=\int dq_2 W_{\mbox{\tiny cl}}((q^i_1,t_1),(q^i_2,t_2))W_{\mbox{\tiny cl}}((q^i_2,t_2),(q^i_3,t_3))
\ee
Since the extremal curves all go through the single point corresponding to $t^r_2$,  we immediately recover  \eqref{equ:record_dec}:
$$W_{\mbox{\tiny cl}}(q^*,q)=W_{\mbox{\tiny cl}}((q^i_1,t_1),(q^i_3,t_3))=W_{\mbox{\tiny cl}}((q^i_1,t_1),(q^{i(r)}_2,t^r_2))W_{\mbox{\tiny cl}}((q^{i(r)}_2,t^r_2),(q_3,t_3)) \simeq W_{\mbox{\tiny cl}}(q^*,q_r)W_{\mbox{\tiny cl}}(q_r,q)
$$

Calculating the probability of $q$ from equation \eqref{equ:record_dec}, we get an equation of conditional  probability, of $q$ on $q_r$, 
\be\label{Prob_records} P(q_r)=P(q|q_r)P(q_r)\ee
Equation \eqref{Prob_records} thus reproduces the Mott bubble equation, \eqref{equ:Mott} for $q_i, q_j$ along the same classical trajectory, and separated by more than $\hbar$.  

Furthermore, it is easy to show that when  $q_1$ is a record of $q_2$ and there is a unique classical path between the two configurations, then the entire path has an ordering of records. Namely,  parametrizing the path, $\gamma(t)$,  such that $\gamma(0)=q_1\,,\gamma(t^*)=q_2$, then $\gamma(t)$ is a record of $\gamma(t\rq{})$ iff $t<t\rq{}$. We call such types of objects, strings of records, and it is through them that we recover a notion of classical time.

 \subsubsection{The recovery of classical Time}
    If records are present, it would make absolute sense for `observers'  in $q$ to attribute some of $q$\rq{}s properties to the `previous existence'  of $q_r$. It is as if configuration $q_r$ had to  \lq{}happen\rq{} in order for $q$ to come into existence. If $q$ has some notion of history, $q_r$ participated in it. 

 When comparing relative amplitudes between possibly finding yourself in configurations $q_1$ or $q_2$, both possessing the same records $q_r$, the amplitude $W(q^*,q_r)$ factors out, becoming irrelevant. This says that we don\rq{}t need to remember  what the origin of the Universe was, when doing experiments in the lab.\footnote{But note that whenever a record exists, the preferred configuration $q^*$ is also a record. In fact, one could have defined it as \emph{the} record, of all of configuration space. Indeed, it does have the properties of being as unstructured as possible, which we would not be amiss in taking to characterize an origin of the Universe. }

I believe that indeed, it is difficult to assign meaning to some future configuration $q$ in the timeless context. Instead, what we do, is \emph{to compare expectations `now', with retrodictions, which are embedded in our records, or memories}.  We compare earlier records with more recent ones. 
When we have a record at $q$,  then $q_r$ itself acquires  meaning. Accordingly, records imbue  $W(q_r,q)$ with stronger epistemological status. 

 This finally gives us back a complete notion of history, which is recovered only in the complete classical limit ($\hbar\rightarrow 0$). 
This retrodiction aspect begs for a Bayesian treatment of probabilities, which works well (but  we leave this analysis for the appendix \ref{sec:Bayes}).

  If there is nothing to empirically distinguish between our normal view of  history (i.e. as having actually happened) on one hand, and the tight correlation between the present and the embedded past on the other, why should we give more credence to the former interpretation? Bayesian analysis can pinpoint no pragmatic distinction, and  I see no reasons for preferences, except psychological ones. 
 
 \subsubsection{Records and conservation of probability}
Now, one of the main questions that started our exploration of theories that are characterized by the timeless transition amplitude, was the difficulty in defining concepts such as conservation of probability for quantum gravity, which has no fixed causal structure. Are we in a better position now?

What we are talking about so far is volume in configuration space. How does that relate to probabilities, of the sort that is conserved? But first, conserved in which \lq{}time\rq{}? In the presence of a standard time parameter, we first distinguish between the total probability $P_t$ at one time, $t$, from $P_{t\rq{}}$ at another, $t\rq{}$. To translate this statement to one that uses only records and configuration space, we want a notion that reproduces this separation. This separation is easily accomplished by first restricting  configurations in $\mathcal{M}_{(r)}$ to subsets, $\mathcal{S}_\alpha$, $\alpha \in \Lambda$, such that there is no pair $\phi^\alpha_i, \phi^\alpha_j\in \mathcal{S}_\alpha$ for which $\phi_i\in \mathcal{M}_{(j)}$. We call these sets, $\mathcal{S}_\alpha$, \textit{screens}.  In other words, in each one of these sets, no configuration is a record of any other configuration. This is taken to say that configurations belonging to a single screen are not ``causally related''. In relativistic terminology --- which can be misleading, since here we are in configuration space, and not in real space(time)--- this would represent events that happen `at the same time' (for some equal time surface). But here this property also holds for Many-Worlds type theories in configuration space, where there is no real time. 

Now, each $\mathcal{S}_\alpha\subset\mathcal{M}_{(r)}$ does not contain redundant records. But there are many redundant records along each extremal trajectory, at least in the no-interference case.  In that simple case, there is precisely one extremal trajectory $\gamma_j$ between $\phi_r$ and each element $\phi^\alpha_j$ of a given screen, which is thus parametrized by the set $J\ni j$. Define a screen $\mathcal{S}_1=\{\phi^1_j=\gamma_j(t^1_j), j\in J\}$, where $t_j^1$ is a given parameter along the $j$-th extremal curve.  We can then find another screen $\mathcal{S}_2=\{\phi^2_j=\gamma_j(t^2_j)\,,\,\, t^2_j> t^1_j, \forall j\}$. In these simple cases, and at least for certain types of action functionals, it can be shown that, for the translationally invariant measure and Born volume, the infinitesimal volumes respect: $V(\mathcal{S}_2)\simeq V(\mathcal{S}_1)$ \cite{QG_deco}. This is as close as we can get to a statement about conservation of probability.

\section{What are we afraid of? The psychological obstacles.}\label{sec:Parfit}
What usually unsettles people -- including me -- about this view is  the damage it does to the idea of a continuous conscious self.  The egalitarian  status of each and all instantaneous configurations of the Universe -- carrying on their backs our own present conscious states --  raises alarms in our heads. Could it be that each instant exists only unto itself, that all our myriad instantaneous states of mind \emph{exist separately}? This  proposal appears to conflict with the construed narrative  of our selves -- of having a continuously evolving and self-determining conscious experience.

  But perhaps,  upon reflection, it shouldn\rq{}t bother us as much as it does. First of all, the so-called Block Time view of the self does not leave us in much better shape in certain respects of this problem. After all, the general relativistic worldline does not imply an \lq{}evolving now\rq{} -- it implies a collection of them, corresponding to the entire worldline. For the (idealized)  worldline of a conscious being,  each element of this collection will have its own, unique, instantaneous experience. 
  
Nonetheless, in at least one respect,  the worldline view still seems to have one advantage over the one presented in this paper.  The  view presented here  still appears more fractured, more disconnected, less linear than the worldline view, even in the classical limit -- for which aspects of a  one-parameter family of configurations becomes embedded in a \lq{}now\rq{}.  The problem is that we start off with a one-parameter family of individual conscious experiences, and, like Zeno, we imagine that an inverse limiting procedure focusing on the \lq{}now\rq{}  will eventually tear one configuration from the \lq{}next\rq{}, leaving us stranded in the \lq{}now\rq{}, separated from the rest of configuration space by an infinitesimal chasm. This is what I mean here by \lq{}solipsism of the instant\rq{}. 
In the first subsection of this section, I  will explain how this intuitive understanding can only find footing in a particular choice of (non-metric) topology for configuration space. That topology  is not  compatible with our starting point of applying differential geometry to configuration space.

  
  \subsection{Zeno's paradox and solipsism of the instant: a matter of topology}\label{sec:zeno}


 It might not seem like it, but the discussion about whether we have a `a collection of individual instants'  as opposed to `a continuous curve of instants'  hinges, albeit disguisedly, on the topology we assume for configuration space. Our modern dismissal of Zeno's paradox relies on the calculus concept of a limit. But in fact, a \emph{limit point} in a topological space first requires the notion of topology:  a limit point of a set $C$ in a topological space $X$ is a point $p\in X$ (not necessarily in $C$) that can be ``approximated" by points of $C$ in the sense that every neighborhood of $p$ with respect to the topology on $X$ also contains a point of $C$ other than $p$ itself. 

In the finest topology -- the discrete topology --  each subset  is declared to be open. On the real line, this would imply that every point is an open set. Let us call an abstract pre-curve in $X$ the image of an injective mapping from  $\mathbb{R}$ (endowed with the usual metric topology) to the set $X$. Thus no pre-curve on  $X$ can be continuous if $X$ is endowed with the finest topology.  Because the mapping is injective, the inverse of each point of its image (which is an open set in the topology of $X$) is a single point in $\mathbb{R}$, which is not an open set in the standard metric topology of $\mathbb{R}$. Likewise,  with the finest topology, Zeno's argument becomes inescapable -- when every point is an open set, there are no limit points and one indeed cannot hop continuously from one point to the next. We are forever stuck \lq{}here\rq{}, wherever here is. 

In my opinion, the idea that Zeno and Parmenides were inductively aiming at was precisely that of a discrete topology, where there is a void between any two given points in the real line. 
If $X$ is taken to be configuration space, this absolute ``solipsism of the instant"  would indeed incur on the conclusions of the Eleatics, and frozen time would necessarily follow.  However, \emph{the finest topology cannot be obtained by inductively refining metric topologies.} 

  With a more appropriate, e.g. metric,  topology,  we can only iteratively get to open neighborhoods of a point, neighborhoods which include a continuous number of other configurations. That means for example that smooth functions on configuration space, like $P(q)$, are too blunt an instrument -- in practice its values cannot be used to distinguish individual points. No matter how accurately we measure things, there will always be open sets whose elements we cannot parse. And so it is with any subjective, empirical notion of \lq{}now\rq{}. 
  

The point being that with an appropriate topology we can have timelessness in a brander version than the Eleatics, even assuming that  reality is entirely contained in configuration space  without any absolute time. With an appropriate coarser (e.g. metric) topology on configuration space, we do not have to worry about a radical ``solipsism of the instant": in  the classical limit there are continuous curves interpolating between a record and a record-holding configuration.   I can safely assume that  there is a \emph{continuous sequence} of configurations connecting me eating that donut this morning to this present moment of reminiscence. 

\subsection{The continuity of the self - Locke, Hume and Parfit}
John Locke considered personal identity (or the self) to be founded on memory, much like my own view here. He says in  \lq\lq{}Of Ideas of Identity and Diversity\rq\rq{}:  
\begin{quote}``This may show us wherein personal identity consists: not in the identity of substance, but [...] in the identity of consciousness. [...] This personality extends itself beyond present existence to what is past, only by consciousness" \end{quote}
David Hume, wrote in \lq\lq{} A Treatise of Human Nature\rq\rq{}  that when we start introspecting, ``we are never intimately conscious of anything but a particular perception; man is a bundle or collection of different perceptions which succeed one another with an inconceivable rapidity and are in perpetual flux and movement.". 

Indeed, the notion of self, and continuity of the self, are elusive upon introspection. I believe, following Locke, that our self is determined biologically by patterns in our neural connections.  Like any other physical structure, under normal time evolution these patterns are subject to change. What we consider to be a `self' or a `personality', is inextricably woven with the notion of continuity of such patterns in (what we perceive as) time. Yes, these patterns may change, but they do so continuously. It is this continuity which allows us to recognize a coherent identity.

 In Reasons and Persons, Derek Parfit puts these intuitions to the test. He asks the reader to imagine entering a \lq\lq{}teletransporter\rq\rq{} a machine that puts you to sleep, then destroys you,  copying the information of your molecular structure and then relaying it to Mars at the speed of light. On Mars, another machine re-creates you, each atom in exactly the same relative position to all the other ones. Parfit poses the question of whether or not the teletransporter is a method of travel -- is the person on Mars the same person as the person who entered the teletransporter on Earth? Certainly, when waking up on Mars, you would feel like being you, you would remember entering the teletransporter in order to travel to Mars, you would also remember eating that donut this morning.

Following this initial operation,  the teletransporter on Earth is modified so as to leave intact the person who enters it. Each replica left on Earth would claim to be you, and also remember entering the teletransporter, and then getting out again, still on Earth. Using thought experiments such as these, Parfit argues that any criteria we attempt to use to determine sameness of personal identity will be lacking. What matters, to Parfit, is simply what he calls ``Relation R": psychological connectedness, including memory, personality, and so on. 

This is also my view, at least intellectually if not intuitively. And it applies to configuration space and the general relativistic worldline in the same way as it does in Parfit's description. In our case there exists a past configuration, \emph{represented} (but not contained) in configuration `now' in the form of a record. This past configuration  has in it neural patterns that bear a strong resemblance  to neural patterns contained in configuration `now'. Crucially, these two configurations are connected by \emph{continuous extremal paths in configuration space},  ensuring that indeed we can act as if they are psychologically connected.  We can --- and should! --- act as if one classically evolved from the other. Indeed, in such cases  our brain states are consistent with  evolved relations between all subsystems in the world  we have  access to. Different sets of records all agree and are compatible with arising from joint evolution. Furthermore, I would have a stronger Relation R with what I associate with future configurations of my (present) neural networks, than to other brain configurations (e.g. associated to other people).  There seems  to be no further reason for this conclusion to upset us, beyond those reasons that  already make us  uncomfortable with Parfit\rq{}s thought experiment.  

\section{Summary and conclusions}

The idea of timelessness is certainly counter-intuitive.  

But our own personal histories can indeed be pieced together from the static landscape of configuration space. Such histories are indiscernible from, but still somehow feel less real than our usual picture of our pasts. Even beyond  the worldline view of the self,  the individual existence of \emph{every} instant still seems to leave holes in the integrity of our life histories. I have argued this feeling is due to our faulty intuitions about the topology of configuration space. 

   Nonetheless, even after ensuring mathematical continuity of our notion of history,   the idea of timelessness and of \emph{all} possible states-of-being threatens the ingrained feeling that we are self-determining; since all these alternatives exist timelessly, how do we determine our future? 
But this is a hollow threat. Forget about timelessness; free will and personal identity are troublesome concepts all on their own, we should not fear infringing their territory.  I like to compare these concepts to mythical animals: Nessie, Bigfoot,  unicorns and the like. They are constructs of our minds, and -- apart from blurry pictures --  shall always  elude close enough inspection.   Crypto-zoologists notwithstanding,  Unicorns are not an endangered species. We need not be overly concerned about encroaching on their natural habitat. 

\subsection{Crippling and rehabilitating Time}

\subsubsection{Crippling Time}
\begin{quote}
\lq\lq{}Time does not exist. There is just the furniture of the world that we call instants of time. Something as final as this should not be seen as unexpected. I see it as the only simple and plausible outcome of the epic struggle between the basic principles of quantum mechanics and general relativity. For the one -- on its standard form at least -- needs a definite time, but the other denies it. How can theories with such diametrically opposed claims coexist peacefully? They are like children squabbling over a toy called time. Isn\rq{}t the most effective way to resolve such squabbles to remove the toy?\rq\rq{}\cite{Julian_End}
\end{quote}

 Loosely following the Eleatic view of the special ontological status of the present, here we have carved Time away from spacetime, being left with timeless configuration space as a result. If Time is the legs  which carries space forward, we might seem to have emerged from this operation with a severely handicapped Universe. 
 
The criticism is to the point. Even if Time does not exist as a separate entity in the Universe, our conception of it needs to be recovered somehow. If there is no specific variable devoted to measuring time, it needs to be recovered from  relational properties  of configurations.  
This essay showed that this can be done.

    \subsubsection{Rehabilitating Time} 

The plan was to recover Time by using  a semi-classical approximation of a fundamentally timeless quantum mechanics theory in configuration space.
There are two meanings of Time which should be discerned. One is merely an ordering of configurations. The other is a a sense of duration. I have not discussed duration in this paper, but only briefly mention it in a paragraph below.

 \subsubsection{Recovering ordering.}
 For the ordering of configurations, mathematically,  before all else we needed to find a way to breath life into the configuration space propagator, $W(q_1,q_2)$, by defining a second configuration other than any present configuration, \lq{}now\rq{}.

This first step was accomplished by defining a preferred configuration $q^*$, playing a role similar to the quantum mechanical \lq{}vacuum\rq{}. It is preferred in that it is  the most structureless point of reduced configuration space, and  it is also the configuration representing the  \lq\lq{} pointiest corner\rq\rq{} of configuration space.  The precise choice of $q^*$ depends both on the topology of the spatial manifold $M$ and on the relational symmetry group at hand. For the case of $M=S^3$ and the symmetry group Diff$(M)\ltimes\mathcal{C}$, $q^*$ is represented by (the orbit of) the round metric on $S^3$. 

With the introduction of the preferred \lq{}in\rq{} configuration, we defined a positive scalar density (a volume form, or probability density) on configuration space, $P(q)\mathcal{D}q$, given by the transition probability from the vacuum to the given configuration. Restrictions of locality (see \cite{Locality_riem}) and factorization properties (required to make $V(q)$ compatible with the later introduction of records), limit our choices, leading us to conjecture that we can recover the Born form, $P(q)=|W(q^*,q)|^2$.

But in any case,  this state of affairs is still not completely satisfactory. The number $|W(q^*,q)|^2$, which makes reference to some quantum mechanical vacuum state, or \lq\lq{} preferred initial configuration\rq\rq{},  is too removed from everyday practice of physics, and it seems to say nothing about why I believe I really had a donut this morning. To get around this, we need the emergence of records which are more local (in configuration space). In the semi-classical path integral representation, there indeed exist such  candidate objects to play this role. We have named these objects  \emph{semi-classical records}  (or just  records here). 

 If $q_r$ is the record possessed by the configuration `now', $q$,  then $W(q_r,q)$ can encode our immediate pasts, through a correlation of amplitudes. \emph{The probability of $q$ becomes the conditional probability of $q$ given $q_r$}, $P(q)=P(q|q_r)P(q_r)$.  Using the notion of records shifts the burden from predictability in physics, to updating of our knowledge. We do physics by comparing our expectations `now'  --  dictated by our records and our theories --  with other properties of the present configuration. 
 From a Bayesian perspective, it proves worthwhile to assume the existence of a history. 
  
 Indeed, in certain circumstances,  all of the configurations between $q_r$ and $q$ will themselves define an ordering of records,  reproducing a notion of granular history. { In such cases,  each previous run of an experiment is encoded in each posterior run. }  The equations emerging for a Bayesian treatment for the fitness of a given theory are identical to the usual, timeful ones, as was shown in \cite{QG_deco}. 
 
 The more records a certain configuration has, the more data one has to test their theories. Consistency of multiple records within a given configuration increases our level of confidence in a theory, and inconsistency decreases it, as expected. The more such types of consistent structures a configuration has, the more we will have perceived time to have passed. In other words, an arrow of time points from $q^*$ -- the most symmetric configuration -- to ones that have consistent records.

 We have gone from a handicapped picture of a Universe that limps, to one that lilts, step by step conducted by the  complex structures present in configuration space.  

\subsection{Consequences for quantum gravity.}
\subsubsection{Relationalism and Laws of the Instant}

 Looking for a theory that will allow a more natural description of unitarity, probability, and superposition,    we are led to require of it no external input other than what is contained in configuration space itself. Interestingly, this demand puts severe restrictions on the types of symmetries that can exist.  Namely, we only allow those symmetries whose action depends solely on the configuration on which it is acting. The ones that obey these restrictions  we call \lq{}Laws of the Instant\rq\rq{}. 
  
  Had we allowed symmetries which are not \lq{}Laws of the Instant\rq{}, the transition amplitude would not have been a well-defined object by itself, and records would not be invariant under such symmetries, thus losing objective meaning. 
 Fortuitously,  we find that the most general Laws of the Instant are those that indeed embody the full gamut of relational symmetries. 
 
I want to repeat once more  that the theory of general relativity does not accommodate all manifestations of spacetime relationalism. In particular, it does not incorporate relationalism of scale: the theory is not conformally invariant (unlike its unitarily-challenged cousin, conformal gravity).  In a 3+1 formulation, general relativity still does not respect the Laws of the Instant; refoliations are not intrinsic to configuration space.  
 
 Spatially relational theories also have better control over questions of unitarity (unlike conformal gravity), and  have the correct number of degrees of freedom (unlike Horava gravity), although we have said  little in these directions here. Moreover,  configuration space only has a  principal fiber bundle structure for symmetry groups which are \lq{}Laws of the Instant\rq, and this is a  useful structure to have for explicitly writing the path integral \cite{Conformal_geodesic}. 
 In fact, as we saw, in the Hamiltonian language, the natural setting for quantum mechanics, it is impossible to implement relativity of simultaneity without also implementing the equations of motion. This is part of the mixing between evolution and symmetries present in GR, in many different guises. 

   Actions which represent completely spatially relational theories would be the ones suggested by the principles expounded on in this paper. The challenge is to find  ones  which are also in accord with experiment. From these principles, the derivation of shape dynamics, with its non-local Hamiltonian and preferred Time, is without a doubt overly contrived, and requires a detour through general relativistic territory. In the future, we need to investigate the phenomenology of  more natural relational models, such as the one given by \eqref{new action1} in the appendix.   

\subsubsection{A note on duration.}
 In a 3+1 description of GR, given an initial and a final Cauchy surfaces and a unique spacetime interpolating between the two, duration along a world-line is read off from the lapse associated to that foliation. The lapse is a Lagrange multiplier, associated to relativity of simultaneity.  In these types of relational configuration space theories, e.g.: \eqref{equ:SD_Hamiltonian} and \eqref{new action1}, there is no inkling of a lapse anywhere to be seen.  For a given  unique extremal field history between $[g_1]$ and $[g_2]$, one must then \emph{define} duration as a local measure of change in the conformal geometry.  In this case, duration along a worldline is completely relational, and does not set a scale by itself, unlike what is the case  in GR.  {Not only is this possible, but it has been done for one theory that is not intrinsically formulated in spacetime.  In \cite{Tim_effective, Tim_duration}, it was shown that weak matter perturbations evolving through standard unitary Hamiltonian evolution in shape dynamics \cite{SD_first, Flavio_tutorial}, will perceive conformal geometry change -- or duration -- in the exact proportion to rebuild an Einstein spacetime (in a particular foliation,  CMC).}

  \subsubsection{What gives, Wheeler\rq{}s quip or superpositions?}
Neither, really. 

Perhaps our shortcomings in the discovery of a viable theory of quantum gravity are telling us that \emph{spacetime} is the obstacle.
Though at first sight we are indeed mutilating the beautiful unity of space and time, this split should not be seen as a step back from Einstein\rq{}s insights. 
    I believe the main insight of general relativity, contained in Wheeler\rq{}s sentence \eqref{sentence:Wheeler}, is about the dynamism of space and time themselves. There is no violence being done to this insight here. 
    
Spatial geometry appears dynamic -- it warps and bends throughout evolution whenever we are in the classical regime. Regarding the dynamism of Time, the notion of  \lq{}duration\rq{} \emph{is} emergent from relational properties of space. Thus duration too, is dynamic and space-dependent. 

  Nonetheless,  all relational properties are encoded in the \emph{static} landscape of configuration space. The point is that this landscape is full of hills and valleys, dictated by the preferred volume form that sits on top of it. From the way that the volume form distributes itself on configuration space, certain classical field histories -- special curves in configuration space -- can give a thorough illusion of change. I have argued that this illusion is indistinguishable from how we perceive motion, history, and time.   
 
  Moreover, with regards to the quantum mechanics adage,  the processes  $W(q_r,q)$ straightforwardly embody ``everything that can happen, does happen".  The concept of superposition of causal structures (or even that of superposition of  geometries), is to be replaced by interference between paths in configuration space. Those same hills and valleys in configuration space that encode classical field histories reveal the valleys and troughs of interference patterns.  A very shallow valley around a point --  for example representing an experimental apparatus \emph{and} a fluorescent dot on a given point on a screen --  indicates the scarcity of observers  sharing that observation.   
  By looking at the processes between records and record-holding configurations, we can straightforwardly make sense of interference, or lack thereof, between (coarse-grained) histories of the Universe.
  
  In all honesty, I don\rq{}t know if formulating a theory in which space and time appear dynamical, and in which we can give precise meaning to superpositions of alternative histories, is enough to quantize gravity. Although the foundations seem solid, the proof is in the pudding, and we must further investigate tests for these ideas. 

But  I also don\rq{}t believe that dropping Time from the picture is abdicating hard-won knowledge about spacetime.   Indeed, we can recover a notion of history,  we can implement strict relationalism, we transfigure the \lq{}measurement problem\rq{}, and  we can make sense of a union of the principles of quantum mechanics and geometrodynamics.\footnote{Perturbative techniques of course still need to be employed, even in the semi-classical limit,  to make sense of the weights $\Delta$ in \eqref{equ:semi_classical}.  This, and other issues to do with renormalizability are left for future study.  } 

  It seems to me that there are many emotions against this resolution, but very few arguments; as I said at the beginning of these conclusions, accepting timelessness is deeply counter-intuitive. But such a resolution would necessarily change only  how we view reality,  while still being capable of fully accounting for how we experience it. The consequences for quantum gravity still need to be unraveled. Even at the classical level, from our search for a natural action embodying timelessness we are led to  \eqref{new action1}, in  appendix \ref{app:examples}; a theory that now needs to be phenomenologically investigated. But this whole approach should be seen as a framework, not as a particular theory. And indeed, in the non-relativistic regime of quantum mechanics, we are not looking for new experiences of reality, but rather for new ways of viewing the ones we can already predict, a new framework  to interpret these experiences with.  This is the hallmark of a philosophical insight,  albeit in the present case one heavily couched on physics.   As  Wittgenstein once said: 
\lq\lq{}Once the new way of thinking has been established, the old problems vanish; indeed they become hard to recapture. For they go with our way of expressing ourselves and, if we clothe ourselves in a new form of expression, the old problems are discarded along with the old garment.\rq\rq{}

\section*{ACKNOWLEDGEMENTS}

I  would like to thank Lee Smolin and Wolfgang Wieland for discussions and help with the writing, and Athmeya Jayaram for introducing me to the work of Derek Parfit. This research was supported  by Perimeter Institute for Theoretical Physics. Research at Perimeter Institute is supported by the Government of Canada through Industry Canada and by the Province of Ontario through the Ministry of Research and Innovation.

\begin{appendix}

\section*{APPENDIX}

\section{Timeless quantum mechanics, the canonical theory}\label{sec:canonical_timeless}

Configuration space, $\mathcal{M}$, is coordinitized by $q^a$, for $a=1,\cdots, n$. An observation yields a complete set of $q^a$, which is called an event. 

 $\Omega=T^*\mathcal{M}$ is the cotangent bundle to configuration space, with coordinates $q^a$ and their momenta $p_a$. The classical dynamics is fully determined once one fixes the Hamiltonian constraint surface in $\Omega$, given by $H=0$, where $H:\Omega\rightarrow \mathbb{R}^k$ is the Hamiltonian of the system.  If $k=1$, then we have a single Hamiltonian constraint, whose action only generates reparametrizations of curves in phase space, if $k>1$, then we have further gauge-invariance. 

A curve $\gamma\in \mathcal{M}$ is a physical motion connecting the events $q^a_1$ and $q^a_2$ if there exists an \emph{unparametrized} curve $\bar\gamma$ in $T^*\mathcal{M}$ such that the following action is extremized:
\be \label{equ:action} S[\bar\gamma]=\int_{\bar\gamma}p_adq^a 
\ee
for curves lying on the constraint surface $H(q^a, p_a)=0$, and are such that $\bar\gamma$'s projection to $\mathcal{M}$ connects $q^a_1$ and $q^a_2$. By parametrizing the curve with a parameter $\tau$, we get the familiar form:
\be\label{equ:parametrized_action}
 S[\bar\gamma]=\int d\tau \left(p_a\dot{q}^a-N_i(\tau)H^i(q^a, p_a)\right)
\ee 
where the $N^i$ are Lagrange multipliers.

One can now define the fundamental transition amplitudes between configuration eigenstates:\footnote{As much as possible, I want to avoid technicalities which won't be required here. Having said this, formally one would have had to define the so-called kinematical Hilbert space $\mathcal{K}$ for the quantum states over $\mathcal{M}$ by using a Gelfand triple over $\mathcal{M}$  with measure $d^dq^a=dq^1\cdots dq^d$, i.e.  $\mathcal{S}\subset\mathcal{K}\subset\mathcal{S}'$. This is not necessary in my case, because we will not require a Hilbert space, as we will see.}
\be
\label{equ:transition}
W(q_1,q_2):=\langle q_1|\hat P|q_2\rangle
\ee where $\hat P$ is the ``evolution operator":
\be\label{equ:projector}
\hat P:=\int d\tau\, e^ {-i\tau \hat H}
\ee
where $\hat H$ is the canonically quantized (with Weyl ordering)  Hamiltonian. Note that since one integrates over all $\tau$, the projector is parametrization independent. To obtain physical states, one need still use the quantization of the constraints, as in \eqref{equ:wdw}:
\be \label{equ:wdw_general}
\hat{H}|\psi\rangle=0
\ee
We will for now implicitly consider the case of a single constraint, $H:\Omega\rightarrow \mathbb{R}$.

Given two regions in configuration space, $R_1, R_2$, we have that the probability of an observation in $R_2$ given an observation in $R_1$ is:
\be
\label{equ:probability}P(R_1,R_2)=\left|\frac{W(R_1,R_2)}{\sqrt{W(R_1,R_1)}\sqrt{W(R_2,R_2)}}\right|^2
\ee
where $${W(R_1,R_2)}=\int_{R_1} dq_1\int_{R_2}dq_2\,  W(q_1,q_2)$$

\subsubsection*{Standard non-relativistic quantum mechanics through deparametrizable systems}
If one can  single out a degree of freedom to parametrize motion on a whole region of configuration space, we can write $q^a=(t,q^i)$, in which case one gets a momenta conjugate to time and writes $H(t,q^i,p_t,p_i)=p_t+H_o(t,q^i,p_i)$.  In this case, by inserting a decomposition of the identity in terms of eigenstates $p_t$ and $E$ of $\hat{p}_t$ and $\hat {H}_o$, one obtains: 
\be
\label{equ:Greens=transition}
W(q^a_1,q^a_2)=W(t_1,q_2^i,t_2,q^i_2)=\int dE\, e^ {-iE(t_1-t_2)}\langle q^i_1|E\rangle \langle E| q^i_2\rangle=G(t_1,q_2^i,t_2,q^i_2)
\ee
where $G(t_1,q_2^i,t_2,q^i_2)$ is the usual transition amplitude in quantum mechanics.\footnote{ One should be careful to note however, that in standard non-relativistic quantum theory, time is not an operator, and thus $\Delta t=0$, i.e. measurements are made at a specific instant. Thus, although at the level of transition amplitudes, equation  \eqref{equ:transition} for deparametrizable systems reproduces $G(t_1,q_2^i,t_2,q^i_2)$, the probabilities for measurements performed with some inaccuracy $\Delta t$ in the time variable need not match. The basic reason is that according to \eqref{equ:probability}, one sums over the transition amplitudes first, and then one takes the squared norm. Thus they are summed interferentially. For standard quantum mechanics with time dependence, one takes the squared norm at each instant and then integrates over the time taken by the measurement.  The temporal resolution $\Delta t$ necessary for a good agreement between the two theories was studied in  \cite{Rovelli_Marolf} for simple systems.  }

\subsubsection*{Gauge transformations}
The presence of  more than one constraint, i.e. $H:\Omega\rightarrow \mathbb{R}^k$, indicates further gauge symmetries of the system. In this case we must impose all of the respective equations \eqref{equ:wdw_general} simultaneously, which generally is difficult. When the the commutation of the constraints form a true  Lie Algebra: 
\be\label{equ:Lie_algebra}
[\hat{H}^i,\hat{H}^j]=f^{ij}_{~~k}\hat{H}^k,
\ee
i.e. when $f^{ij}_{~~k}$ have no phase space-dependence, different methods can be used to find the projection onto the physical states, the most straightforward of which is called `group-averaging' \cite{Marolf}. This consists in integrating over the group:
$$|\Psi\rangle=\int_G d\mu(U) \hat {U} |\psi\rangle
$$ where $d\mu$ is the Haar measure, which is translation invariant. In the more general case, this technique will in general incur in anomalies.\footnote{In the more general case one should use BRST techniques, but in the Lie groupoid case -- i.e. when the structure constants depend on the fields, $f^{ij}_{~~k}(q,p)$ -- is still much more problematic, as one has a BRST charge that is not (usually) of rank 1 in ghost momenta \cite{Henneaux_Teitelboim}.}

\section{Relational symmetries and laws of the instant}\label{sec:LOTI}
In the case at hand, suppose that the transformations in phase space are given by a Hamiltonian vector field, associated to a  smeared functional $F[g,\pi,\eta]$, polynomial in its variables. For  this to have an action on configuration space that is independent of the momenta, $F[g,\pi,\eta]$ must be linear in the momenta. This already severely restricts the forms of the functional to 
$$F[g,\pi,\eta]=\int F_1(g,\eta)_{ab}\pi^{ab}$$
A Poisson bracket here results in
$$
\{F[g,\pi,\eta_1],F[g,\pi,\eta_2]\}=\int d^ 3x  \left(\frac{\delta F_1(g,\eta_1)_{ab}}{\delta g_{cd}}\pi^{ab}F_1(g,\eta_2)_{cd}-\frac{\delta F_1(g,\eta_2)_{ab}}{\delta g_{cd}}\pi^{ab}F_1(g,\eta_1)_{cd}\right)
$$
where $F_1(g,\eta)_{ab}$ must be a covariant tensor of rank two.

If $F_1$ has no derivatives of the metric, it will straightforwardly commute. But with no derivatives the only objects we can form are: 
$$ F_1(g,\eta)_{ab}=\eta g_{ab}\, ~,~\mbox{and} \, ~ F_1(g,\eta)_{ab}=\eta^ {ab}g_{ab}
$$
In the first case, these are just conformal transformations, in the second, they would imply that $\pi^{ab}=0$, a constraint killing any possibility of dynamics, which is still consistent (also consistent with the strictly Eleatic  Universe).

The point now is to show that only with one derivative -- which implies  a Lie derivative for a covariant object -- they still weakly commute. With more derivatives of the metric, the conjecture is that one does not close the algebra. For example,  
$$ F_1(g,\eta)_{ab}= \eta(\alpha R_{ab}+\beta R g_{ab})
$$
it is straightforward but tedious to show that the algebra does not close for any values of $\alpha$ and $\beta$. If one instead chose a term of the form $\beta R\eta_{ab}$ one can show that the rank of this constraint is not constant along phase space. Furthermore, it implies that almost everywhere $\pi^{ab}=0$, as before.  The conjecture is that these conclusions hold order by order in number of derivatives of the metric.

\subsection{Configuration space metrics}

\subsubsection{Superspace}
One nice thing about the space of metrics is that itself comes with a supermetric, defined, for $v\,,w\in T_g\mathcal{M}$, at the base point $g_{ab}\in \mathcal{M}$ by:\footnote{In fact, it comes with a one-parameter family of supermetrics, where we substitute $g^ {ac}g^{bd}\rightarrow g^ {ac}g^{bd}+\lambda g^ {ab}g^{cd}$. This supermetric however is only positive for $\lambda>-1/3$. Note also that we are not symmetrizing the DeWitt supermetric because we are assuming that it is acting on tangent vectors of $\mathcal{M}$, which are already symmetric.}
\be\label{equ:supermetric}
\langle v\,, w\rangle_g=\int d^3x\sqrt{g}\, g^ {ac}g^{bd} v_{ab}w_{cd}
\ee
This supermetric induces a metric topology on $\mathcal{M}$.\footnote{More formally, one would work with what is called a weak Whitney topology, which roughly is a norm on the jet bundles of the sections. We will ignore these more formal aspects here. } Furthermore, the inner product \eqref{equ:supermetric} is invariant wrt to diffeomorphisms acting through pull-back. That is, the directions along the diffeomorphism orbits in $\mathcal{M}$ are Killing wrt the metric \eqref{equ:supermetric} (see \cite{Ebin}).

\subsubsection{Conformal Superspace}
However, the supermetric  \eqref{equ:supermetric} is not invariant wrt conformal transformations, because of the presence of $\sqrt{g}$. Thus we construct:
\be\label{equ:conformal_supermetric}
\langle v\,, w\rangle_g=\int d^3x\sqrt{g}\,\sqrt{C^{ef}C_{ef}}\, g^ {ac}g^{bd} v_{ab}w_{cd}
\ee
where $C_{ab}$ is the Cotton-York tensor, which is both traceless and transverse, and has the correct conformal weight for the inner product to be conformally invariant. 

\subsubsection{Examples of relational conformal geoemetrodynamical theories}\label{app:examples}
Let me exemplify the constructions above with two different actions. The first is that of shape dynamics, and is given in Hamiltonian form by 
\be
\label{equ:SD_Hamiltonian} H_{SD}=\int \sqrt g\left( (e^{6\phi_o}-\rho\pi^{ab}g_{ab}-\pi^{ab}\mathcal{L}_\xi g_{ab}\right)
\ee
where $\mathcal{L}_\xi $ denotes the Lie derivative, and $\phi_o$ is defined implicitly from the modified Lichnerowicz-York equation \cite{York}:
 \be
 \label{equ:LY}
e^{-6 \phi} \frac{ \pi_{ab} \pi^{ab}}{\sqrt{g}} + \sqrt{g} \left(e^{6 \phi} (-(t^2/6) + 2\Lambda) + 
     e^{2 \phi} (-R + 8 (\nabla_a\phi\nabla^a\phi+\nabla^2\phi))\right)=0
  \ee 

The problem with \eqref{equ:SD_Hamiltonian} is that it can be put in this form only when it is deparametrizable (see \cite{Flavio_tutorial}). Otherwise, one must use not the full group of scale transformations as symmetries, but only those that preserve the \emph{total} volume of space. I.e. instead of $\rho\pi^{ab}g_{ab}$ one must use $(\rho-\langle\rho\rangle)\pi^{ab}g_{ab}$ where 
$$\langle\rho\rangle=\frac{\int\sqrt{g}\rho}{\int\sqrt g} $$
This is a non-local restriction, and is hard to make sense of from a purely relational manner.

The second action is given in its Lagrangian Jacobi form (see \cite{York, Conformal_geodesic}):
We have the reparametrization and conformal-diffeomorphism invariant geodesic action: 
\be\label{new action1} S= \int dt\sqrt{\int_M
\sqrt{g}d^3x \sqrt{C^{ab}C_{ab}}\left(\dot g^{cd}-(\mathcal{L}_\xi g)^{cd}-\rho g^{cd}\right)\left(\dot g_{cd}-(\mathcal{L}_\xi g)_{cd}-\rho g_{cd}\right))}
  \ee
  where $\xi^a$ and $\rho$ are the Lagrange multipliers corresponding to diffeomorphisms and conformal transformations, respectively. 
  
 To stress, this is a fully conformal diffeomorphism invariant action with the same physical degrees of freedom as general relativity, but
which does {\it not} have  local refoliation invariance, only a global reparametrization one.  Equation  \eqref{new action1} is furthermore a purely geodesic-type action in Riem, with just one global lapse and thus one global notion of time, as such it also possesses inherent value in a relationalist setting.
Classical solutions are one-parameter collections of conformal geometries, which extremize the total length according to a given supermetric \eqref{equ:conformal_supermetric}. Although this theory is completely relational --  it is not yet clear whether it will reproduce standard tests of general relativity, and more analysis is required.

\section{Bayesian analysis}\label{sec:Bayes}

Let us call `an observation' $E$ a property of correlations within a configuration. We call the manifold $\mathcal{M}_{(E_o)}$ the manifold which has records of observation $E_o$. 
 Given a theory  $\mathbb{T}_i$ (where $i$ indexes the theory we are discussing), the probability of observation $E_1$ is given by the relative volume of observers: 
  \be \label{equ:volume_observer}P(E|\mathbb{T}_i)=\frac{P_i(E_1)}{P_i(\mathcal{M}_{(E_o)})}:=\frac{\int_{E_1}F_i(\phi)\mathcal{D}\phi}{\int_{\mathcal{M}_{(E_o)}}F_i(\phi)\mathcal{D}\phi}\ee

    In Bayesian analysis we want to judge the ability of the theories to explain a given distribution of the observation given the theory 
This number is called the \emph{`likelihood' of the theory} $\mathbb{T}_i$ given the observation of $E$.   
   We want to compare the chances that we will find ourselves correlated with a  ``new" observation $E_1$, according to the two distinct theories above. Assuming  that  $F$ factorizes according to \eqref{equ:factorization_density}, 
$$\int_{E_1}F_i(\phi)\mathcal{D}\phi=F_i(K(\phi^*,E_o))\int_{E_1}F_i(K(E_o,\phi))\mathcal{D}\phi
$$  
   and using Bayes rule,  we can determine the posterior probability of the theory given the records $E_o$ and observation $E_1$:
\be\label{equ:bayes} P(\mathbb{T}_i|E_1)\approx  \frac{P(\mathbb{T}_i)P(E_1| \mathbb{T}_i,E_o)}{P(E_1)}
\ee 
where 
 $P(E_1| \mathbb{T}_i,E_o)$ is obtained from \eqref{equ:volume_observer} by the replacement $F_i(\phi)\rightarrow F_i(K(E_o,\phi))$. It should be interpreted as the probability that you would find $E_1$ \emph{if} $T_i$ is correct, and you have already \lq{}observed\rq{} $E_o$ (it is a record). 
   For configurations that have many records of properties of $E$ we can -- by just looking at these records  -- update the prior of the next (or present) such observation $E_n$. This is the standard way in which we test our theories, nothing needs to change.

\end{appendix}



\end{document}